\documentclass[reprint,nofootinbib,amsmath,amssymb, amsfonts,floatfix,aps,prd]{revtex4-2}
\usepackage[utf8]{inputenc}
\usepackage{isotope}
\usepackage{textcomp}
\usepackage{gensymb}
\usepackage{booktabs}
\usepackage{graphicx,subfigure}
\usepackage{multirow}
\usepackage{float}
\usepackage{hyperref}

\newcommand{\nuwro}{\textsc{NuWro}}
\newcommand{\gibuu}{\textsc{g}i\textsc{buu}}
\newcommand{\MB}{MicroBooNE}
\newcommand{\MBC}{MicroBooNE Collaboration}
\newcommand{\p}{CC$1p0\pi$}
\newcommand{\pp}{CC$2p0\pi$}
\newcommand{\thetaLR}{\ensuremath{\theta_{pp}}}
\newcommand{\cosThetaLR}{\ensuremath{\cos\theta_{pp}}}
\newcommand{\dsdcosThetaLR}{\ensuremath{d\sigma/d\cos\theta_{pp}}}
\newcommand{\thetaMuPP}{\ensuremath{\theta_{\mu2p}}}
\newcommand{\cosThetaMuPP}{\ensuremath{\cos\theta_{\mu2p}}}
\newcommand{\dsdcosThetaMuPP}{\ensuremath{d\sigma/d\cos\theta_{\mu2p}}}
\newcommand{\dalpha}{\ensuremath{\delta\alpha_T}}
\newcommand{\dsdalpha}{\ensuremath{d\sigma/d\delta\alpha_T}}
\newcommand{\dpT}{\ensuremath{\delta p_T}}
\newcommand{\dsdpT}{\ensuremath{d\sigma/d\delta p_T}}
\newcommand{\dsdpTx}{\ensuremath{d\sigma/d\delta p_{T,x}}}
\newcommand{\etal}{{\it et al.}}
\newcommand{\tk}{\ensuremath{t_\textrm{kin}}}

\newcommand{\ve}[1]{\ensuremath{\mathbf{#1}}}
\newcommand{\n}[1]{\ensuremath{|\mathbf{#1}|}}

\begin{document}
\title{Spectral function approach in NuWro:\texorpdfstring{\\}{} modeling of multinucleon final states in quasielastic scattering}

\begin{abstract}
Neutrino-oscillation experiments performed in the few-GeV energy region create an urgent demand for a significant improvement in the accuracy of modeling neutrino interactions with atomic nuclei. Here, we report an updated implementation of the spectral function approach in the \nuwro{} Monte Carlo generator, which consistently treats multinucleon final states in quasielastic scattering at the inclusive and exclusive levels. After validating its accuracy against inclusive electron-scattering data, we compare its predictions to various neutrino cross sections from \MB{}. We find that with the multinucleon contribution, these data are reproduced with values of $\chi^2$ per degree of freedom of 0.8--2.5, compared to 2.3--7.3 without it.
\end{abstract}

\author{Artur M. Ankowski}
\altaffiliation{Current affiliation: Marian Smoluchowski Institute of Physics, Faculty of Physics, Astronomy, and Applied Computer Science, Jagiellonian University, Ul. Prof. Stanis{\l}awa {\L}ojasiewicza 11, 30-348 Krak{\'o}w, Poland}
\email{artur.ankowski@uj.edu.pl}
\author{Rwik Dharmapal Banerjee}
\author{Jan T. Sobczyk}
\author{\\Jos\'e L. Bonilla}
\author{Krzysztof M. Graczyk}
\author{Beata E. Kowal}
\author{Hemant Prasad}

\affiliation{Institute of Theoretical Physics, University of Wroc\l aw, plac Maxa Borna 9,
50-204, Wroc\l aw, Poland}

\date{\today}

\maketitle

\section{Introduction}
The stringent precision goals of the next generation of accelerator-based neutrino-oscillation experiments~\cite{Hyper-KamiokandeProto-:2015xww,DUNE:2020lwj} require significant progress in the accuracy of our estimates of the cross sections for neutrino interactions with nuclear targets, in particular oxygen and argon. Thanks to increasingly high event statistics, neutrino cross sections are becoming the main source of uncertainties already in ongoing experiments~\cite{NOvA:2021nfi,T2K:2023smv}. In particular, the SBND experiment is expected to record nearly 10 million neutrino interactions with argon within 3 years~\cite{SBND:2025lha}.

In this article, we discuss an improved implementation of the spectral function (SF) approach in the \nuwro{} Monte Carlo (MC) generator, frequently employed by the neutrino community. Going beyond the plane-wave impulse approximation (PWIA), we account for the effects of final-state interactions (FSI) on quasielastic (QE) scattering for targets ranging from carbon to iron. Having in mind applications in the short-baseline program at Fermilab~\cite{MicroBooNE:2015bmn}---comprising SBND, \MB{}, and ICARUS---we pay special attention to the argon target, for which we also model correlated nucleons, stemming from the presence of short-range correlations in the nuclear ground state.  We consistently treat inclusive and exclusive processes, focusing on charged-current (CC) cross sections.

To test the accuracy of the implemented model, we perform comparisons with electron-scattering data. Then, we analyze various differential \pp{} and \p{} cross sections reported from \MB{}~\cite{MicroBooNE:2022emb,MicroBooNE:2023tzj}, for two- or one-proton final states without pions.


While this is the first effort in the literature to discuss the \MB{} \pp{} cross sections~\cite{MicroBooNE:2022emb}, calculations of the \p{} cross sections~\cite{MicroBooNE:2023tzj} were reported before.

Nikolakopoulos \etal{}~\cite{Nikolakopoulos:2024mjj} obtained the QE contributions to the \p{} cross sections in the PWIA and in the distorted-wave impulse approximation (DWIA), and studied how these results are modified by intranuclear cascades implemented in different MC generators, including \nuwro{}. As their calculations were found to generally underestimate the experimental data, the authors of Ref.~\cite{Nikolakopoulos:2024mjj} concluded that two-nucleon knockout contributions play an important role, and that the interference between one- and two-body currents could also yield the missing strength.

McKean \etal{}~\cite{McKean:2025khb} reached similar conclusions based on their implementation of a DWIA approach in the {\sc neut} MC generator. In addition, they suggested that the axial form factor's deviations from the dipole behavior may also contribute to the observed deficit of the \p{} cross section in the region dominated by QE interactions.

Filali \etal{}~\cite{Filali:2024vpy} performed a~joint analysis of the differential cross sections as a~function of transverse kinematic imbalance variables extracted by various experiments, including \MB{}. The framework employed by the authors of Ref.~\cite{Filali:2024vpy} was based on {\sc neut}, combined with \nuwro{} 19.02 to model CC QE interactions in argon. In the context of \MB{}, Filali \etal{} found that \nuwro{} underestimated the effects of the intranuclear cascade. Follow-up results obtained with different generators were summarized in a~recent review~\cite{Dolan:2026nlr}.

A~similar analysis was also performed as part of a~broader study by Yan \etal~\cite{Yan:2025aau}, mainly focused on pion production in neutrino experiments. The authors of Ref.~\cite{Yan:2025aau} employed version 25 of the \gibuu{} MC generator, which obtains two-particle--two-hole contributions to neutrino structure functions from those for electrons, and found very good agreement with the considered \MB{} \p{} cross section.

An analysis by Isaacson \etal{}~\cite{Isaacson:2025cnk} compared predictions of the {\textsc{achilles}} MC generator---based on the SF approach and accounting for QE interactions and single-pion production---with various electron- and neutrino-scattering data, including the double differential \p{} cross sections from \MB{}. As the shape of the experimental data was well reproduced, but not the normalization, Isaacson \etal{} reached similar conclusions to those in Ref.~\cite{McKean:2025khb}.

Liu \etal{}~\cite{Liu:2026wlw} benchmarked a variety of configurations of the {\textsc{genie}} MC generator against the single and double differential \p{} cross sections from \MB{}, comparing different nuclear models, intranuclear cascades, and two parametrizations of the axial form factor. They found that the axial form factor is the most decisive ingredient in reproducing the data, with an averaged lattice QCD result outperforming the fit to historical neutrino-deuteron data.

This paper is organized as follows. In Sec.~\ref{sec:NuWro}, we briefly review the new features in the latest version of \nuwro{}, 25.11, and describe in detail how we implement FSI effects and the correlated SF. We present and discuss our results for electron and neutrino scattering in Sec.~\ref{sec:results}. Finally, in Sec.~\ref{sec:summary} we summarize the article.

\section{\nuwro{} updates}\label{sec:NuWro}
Developed primarily by the neutrino theory group at the University of Wroc{\l}aw, \nuwro{}~\cite{NuWro} is a~MC generator of lepton-nucleus events. Its principal applications are simulations for accelerator-based neutrino experiments in the few-GeV energy region. As \nuwro{} has been described in detail in two recent articles~\cite{Banerjee:2023hub,Prasad:2024gnv}, here we only summarize its salient features pertaining to this analysis.

In the QE channel, \nuwro{} 25.11 offers an option to use the SF approach---both for electron and neutrino scattering---to model nuclear effects in carbon, oxygen, argon, and iron. For argon, the implemented SFs are based on the results of the JLab experiment E12-14-012~\cite{JeffersonLabHallA:2022cit,JeffersonLabHallA:2022ljj}, as reported in Ref.~\cite{Banerjee:2023hub}. For other targets, \nuwro{} employs the SFs of Refs.~\cite{Benhar:1994hw,Benhar:2005dj}, which combine the shell-structure determined in coincidence electron-scattering experiments in Saclay~\cite{Mougey:1976sc,Bernheim:1981si}, with the results of theoretical calculations for infinite nuclear matter at different densities~\cite{Benhar:1989aw}. In addition, for carbon we offer \nuwro{} users an option to employ the new carbon SF~\cite{Ankowski:2024ntv}, which incorporates the experimental input from the high-resolution NIKHEF experiment~\cite{VanDerSteenhoven:1988qa} and consistently combines it with that from the broad-coverage Saclay measurement~\cite{Mougey:1976sc}.

Other interaction channels---available only in the neutrino mode---are simulated in \nuwro{} using the local Fermi gas model. In particular, since the 25.03 update, meson-exchange currents (MEC) can be accounted for in the Valencia 2020 model~\cite{Sobczyk:2020dkn}, and single pion production can be described in the Ghent hybrid model~\cite{Yan:2024kkg}. A noteworthy feature of the Valencia 2020 model, retained in its implementation~\cite{Prasad:2024gnv}, is that it provides predictions for isospin compositions and kinematic distributions of two-nucleon final states. The Ghent hybrid model brought to \nuwro{} an improved description of the second resonance region, interference with the nonresonant background, and a~Regge description at high hadronic masses.

When a lepton interacts with a nucleon bound in a~nuclear target, energy exchanges between the struck nucleon and the spectator system complicate energy conservation at the interaction vertex. These exchanges result in (i) a broadening of the double differential cross section, and (ii) a shift in its position, as a consequence of a modification of the final nucleon's energy spectrum. In what follows, we describe our implementation of these FSI effects in \nuwro{}'s SF approach.

\subsection{Final-state interactions}
To account for the FSI effects in QE lepton scattering off nuclear targets, we implement in \nuwro{} 25.11 the convolution scheme~\cite{Benhar:1991af,Benhar:2006wy,Benhar:2013dq}. This approach involves integrating the PWIA prediction with a folding function that describes the broadening of the cross section induced by interactions between the struck nucleon and the spectator system,
\begin{equation}\label{eq:xsec_FSI}
\frac{d\sigma^\textrm{FSI}}{d\omega d\Omega }= \int d\omega'f_{\ve{q}}(\omega-\omega')\frac{d\sigma^\textrm{PWIA}}{d\omega' d\Omega},
\end{equation}
where $\omega$ is the energy transfer, and $\Omega$ denotes the solid angle specifying the direction of the outgoing lepton.

Since the fraction of nucleons that do not initiate an intranuclear cascade is given by $\sqrt{T_A}$, $T_A$ being the nuclear transparency, the folding function can conveniently be cast in the form
\begin{equation}\label{eq:FF}
f_{\ve{q}}(\omega)=\delta(\omega)\sqrt{T_A}+\big(1-\sqrt{T_A}\big)F_{\ve{q}}(\omega),
\end{equation}
with the finite-width function $F_{\ve{q}}(\omega)$ that induces the broadening effect~\cite{Benhar:2006wy}.

As in Ref.~\cite{Ankowski:2014yfa}, we neglect the $\n q$ dependence of $F_{\ve{q}}(\omega)$ and use its distribution calculated at momentum transfer $\n q = 1$ GeV. This approximation appears to be well justified, considering both the weak dependence of $F_{\ve{q}}(\omega)$ on momentum transfer at large $\n q$ shown in Ref.~\cite{Benhar:2006wy}, and the very good agreement between the calculated cross sections and the available data for electron-scattering on carbon over a broad kinematic region observed in Ref.~\cite{Ankowski:2014yfa}.

We implement the broadening effect by changing the effective energy transfer to the struck nucleon by a~random contribution that follows the $F_{\ve{q}}(\omega)$ distribution. This modification only affects those events in which the struck nucleons initiate intranuclear cascades (``nontransparent events''), and is meant to incorporate mechanisms of interaction beyond the classical treatment.

\nuwro{}'s intranuclear cascade~\cite{Golan:2012wx,Niewczas:2019fro,Dytman:2021ohr} is modified to ensure consistency with the inclusive cross-section calculation: an event treated as nontransparent at the primary vertex involves one or more intranuclear collisions, whereas in an event assumed to be transparent at the primary vertex, the struck nucleon passes undisturbed through the nuclear matter.

Additionally, we alter the density distributions underlying the cascade. Previously, they were approximated by the measured charge densities, $\rho_\textrm{ch}(r)$, parametrized in Ref.~\cite{DeVries:1987atn}. In this analysis, we employ the point densities, $\rho(r)$, unfolded from the charge densities according to Ref.~\cite{Kelly:2002if}.

We do not change other aspects of the cascade, such as the momentum distributions, which still rely on the local Fermi gas model~\cite{Banerjee:2023hub}, and all hadrons propagate as on-shell particles~\cite{Golan:2012wx,Niewczas:2019fro,Dytman:2021ohr}. In channels other than QE scattering, the cascade is the sole FSI effect accounted for in our analysis, and the fraction of transparent events is given by $T_A$.

\begin{figure}
\centering
    \includegraphics[width=0.95\columnwidth]{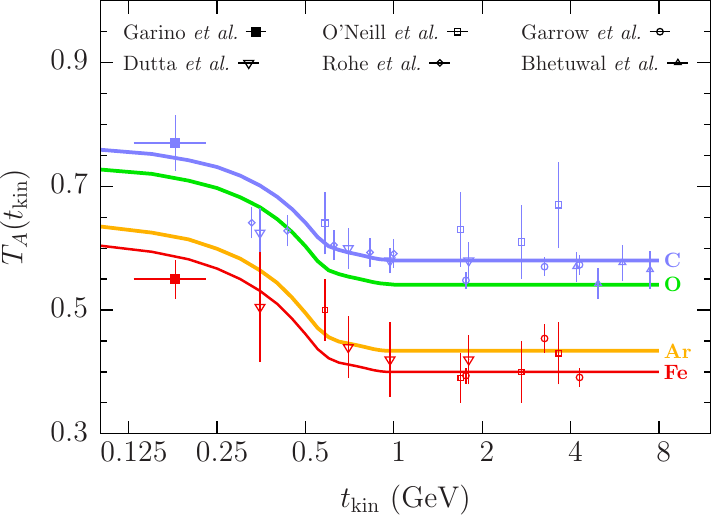}
\caption{\label{fig:transparency}Nuclear transparencies for carbon, oxygen, argon, and iron, implemented in \nuwro{}, are presented as a function of nucleon's kinetic energy on a~base-2 logarithmic scale. The theoretical calculations~\cite{Benhar:2003ka,E97-006:2005jlg} are compared to the measurements reported in Refs.~\cite{Garino:1992ca,ONeill:1994znv,Garrow:2001di,JLabE91013:2003gdp,E97-006:2005jlg,HallC:2022qlb}.
}
\end{figure}

In Fig.~\ref{fig:transparency}, we present the nuclear transparencies implemented in the SF approach in \nuwro{}, and compare them to the experimental results collected from Refs.~\cite{Garino:1992ca,ONeill:1994znv,Garrow:2001di,JLabE91013:2003gdp,E97-006:2005jlg,HallC:2022qlb}. We extend the results for carbon and iron calculated in Refs.~\cite{Benhar:2003ka,E97-006:2005jlg} to oxygen and argon, by assuming that
\[
T_A\propto A^\alpha
\]
where $\alpha =\alpha(\tk{})$ exhibits a dependence on the nucleon's kinetic energy $\tk{}$. We determine it by requiring that the reduction of the nuclear transparency between carbon $(A=12)$ and iron $(A=56)$ is reproduced by construction.

To relate measured $(e,e'p)$ cross sections to nuclear transparencies, experiments typically use MC simulations~\cite{ONeill:1994znv,Garrow:2001di,JLabE91013:2003gdp,E97-006:2005jlg,HallC:2022qlb}. Such an approach leads to model dependence that may be difficult to quantify. It is therefore noteworthy that in the experiment of Garino \etal{}~\cite{Garino:1992ca}, inclusive $(e,e')$ cross sections were used instead to find the normalization. Because Garino \etal{} did not perform a~measurement for iron, in Fig.~\ref{fig:transparency} we present the data point for \isotope[58]{Ni}, which we estimate to differ from \isotope[56]{Fe} by 0.5\%, corresponding to the line width in the figure.

Our tests show a~good agreement between the calculations~\cite{Benhar:2003ka,E97-006:2005jlg} and the measurements~\cite{Garino:1992ca,ONeill:1994znv,Garrow:2001di,JLabE91013:2003gdp,E97-006:2005jlg,HallC:2022qlb} for both carbon, $\chi^2=1.23$ (25.90/21), and iron, $\chi^2=0.89$ (10.70/12). Should the normalization of the theoretical prediction be treated as a~free parameter, the best-fit results would correspond to the nuclear transparency reduced by 2.3\% for carbon, $\chi^2=0.98$ (19.58/20), and increased by 0.5\% for iron, $\chi^2=0.97$ (10.64/11). Based on these findings, we estimate the uncertainty of the nuclear transparency for argon to be 3\%.

In addition to the broadening of the cross section, FSI also induce a modification of the energy spectrum of the final-state nucleon. Following Ref.~\cite{Ankowski:2014yfa}, we account for this effect by including the real part of the optical potential, $U_V=U_V(\tk{})$, in the argument of the folding function,
\begin{equation}\label{eq:rOP}
f_{\ve{q}}(\omega-\omega')\to f_{\ve{q}}(\omega-\omega'-U_V).
\end{equation}

We employ the proton optical potentials determined by Cooper \etal{}~\cite{Cooper:1993nx}. Being based on Dirac phenomenology, these complex potentials are expressed in terms of their scalar and vector components. Their dependence on the kinetic energy and the radial coordinate, $r$, is obtained from fits of the scattering solutions of the Dirac equation to the available body of measurements of
elastic cross sections, analyzing powers, and spin rotation functions for proton scattering off various nuclear targets, including carbon, oxygen, calcium, and iron.

\begin{figure}
\centering
    \includegraphics[width=0.95\columnwidth]{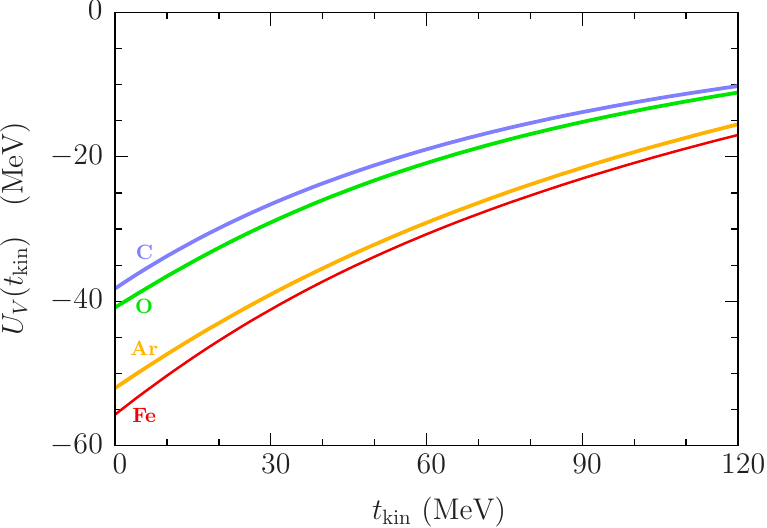}
\caption{\label{fig:potentials}Real parts of the optical potentials for carbon, oxygen, argon, and iron, implemented in \nuwro{}, are presented as a function of the proton's kinetic energy. They are obtained from the Dirac phenomenological fits of Cooper \etal{}~\cite{Cooper:1993nx}.
}
\end{figure}

Using the point densities of the considered targets, we perform a $\rho(r)$-weighted average of the real parts of the optical potentials for proton scattering~\cite{Cooper:1993nx}, as described in Ref.~\cite{Ankowski:2014yfa}, and arrive at the results presented in Fig.~\ref{fig:potentials}. At low kinetic energies, FSI sizably modifies the struck proton's spectrum. As this kinematics is relevant to QE scattering, this effect turns out to be essential to reproduce experimental results, as we will discuss in Sec.~\ref{sec:results}.

We assume that the real parts of the optical potentials for neutrons only differ from those for protons due to the $\rho(r)$-weighted average values of the Coulomb potentials, collected in Table~\ref{tab:VC}.

\begin{table}[t]
\caption{\label{tab:VC} Average Coulomb energies.}
\begin{tabular*}{\linewidth}{@{\extracolsep{\fill}} c c c c c}
    \toprule
    target    & \isotope[12][6]{C} & \isotope[16][8]{O} & \isotope[40][18]{Ar} & \isotope[56][26]{Fe} \\
    $V_C$ (MeV)              & 3.5 & 4.2 & 7.3 & 9.6\\
    \bottomrule
\end{tabular*}
\end{table}

It is important to note that in our approach, both $T_A=T_A(\tk)$ and $U_V=U_V(\tk)$ are evaluated at an approximate value of the nucleon's kinetic energy, obtained for scattering on a nucleon at rest. For a massless probe, it simplifies to
\begin{equation}\label{eq:t}
\tk=\frac{E_{\mathbf k}^2(1 - \cos\theta)}{ M + E_{\mathbf k}(1 - \cos\theta)},
\end{equation}
where $E_{\mathbf k}$ is the probe's energy, $\theta$ denotes the scattering angle of the outgoing lepton, and $M$ is the nucleon mass. This feature stems from the fact that in Eq.~\eqref{eq:xsec_FSI}, the nucleon's momentum is integrated out. As in Ref.~\cite{Ankowski:2014yfa}, the \nuwro{} implementation does not neglect the mass of the produced charged lepton in CC QE neutrino interactions. Within an event, the value of the potential is fixed for all nucleons taking part in secondary interactions.

For the final comments of this subsection, we would like to observe that so far no measurements of the nuclear transparencies of oxygen and argon have been performed, and little is known about the accuracy of the optical potential of argon in the broad kinematic range of interest to the neutrino-oscillation programs. In addition, as optical potentials are determined from nucleon elastic scattering data, they may not be able to provide an accurate description of FSI at kinematic setting where QE scattering and resonance production significantly overlap. We will return to this issue in Sec.~\ref{sec:results}.

\subsection{Correlated spectral functions}
For the argon target, the SFs have recently been obtained by the coincidence electron-scat{\-}ter{\-}ing experiment E12-14-012~\cite{JeffersonLabHallA:2022cit,JeffersonLabHallA:2022ljj}, conducted in Hall~A of Thomas Jefferson National Laboratory, in Newport News, Virginia. In Ref.~\cite{Banerjee:2023hub}, we have reported their implementation in \nuwro{}.

In this analysis, we significantly improve upon their treatment in \nuwro{} by separating the mean-field (MF) and correlated parts of the SFs, and modeling correlated nucleons within the model of Ref.~\cite{CiofidegliAtti:1995qe}, underlying the simulations of the E12-14-012 experiment.

Contributing $\sim$80\% of the SF's strength, the MF part describes the distribution of individual nucleons in the shell-model states. When a MF nucleon takes part in the interaction, no other nucleons are involved in the primary vertex.

The correlated part, constituting $\sim$20\% of the total strength, models the distribution of nucleons taking part in short-range interactions. When the struck nucleon belongs to the correlated contribution, a~correlated nucleon is associated with the primary vertex.

To describe short-range interactions, we employ the model of Ref.~\cite{CiofidegliAtti:1995qe}, based on the assumption that their dynamics is largely decoupled from the $(A-2)$-nucleon system. As they overwhelmingly produce quasi-deuteron pairs~\cite{JeffersonLabHallA:2020wrr}, we disregard the $nn$ and $pp$ contributions.

The correlated part of the spectral function can then be expressed as
\begin{equation}\label{eq:corrSF}
\begin{split}
P_\text{corr}(\ve p,E) &=\int d^3h\,\delta\left(E - E_\text{thr}-T_{A-1}\right)\\
&\qquad\times n^{pn}_\text{cm}(|\ve p +\ve h|)n_\text{rel}^{pn}\left(\Big|\frac{\ve p - \ve h}2\Big|\right),
\end{split}
\end{equation}
where $\ve p$ and $\ve h$ are the initial momenta of the struck neutron and the correlated proton, the threshold energy $E_\text{thr}=25.20$ MeV~\cite{JeffersonLabHallA:2022ljj}, and $T_{A-1}$ is the kinetic energy of the relative motion of the correlated proton and the $(A-2)$-nucleon system,
\[
T_{A-1}=\frac{1}{2\mathcal M}\left(\frac{\ve p}{A-1}+\ve h\right)^2.
\]
with $\mathcal M= M(A-2)/(A-1)$, $M$ being the nucleon mass.

The center-of-mass momentum of the $pn$ pair is assumed to follow the Gaussian distribution,
\begin{equation}
n^{pn}_\text{cm}(p)=\left(\frac{\alpha_\text{cm}}{\pi}\right)^{3/2}\exp(-\alpha_\text{cm} p^2),
\end{equation}
with $\alpha_\text{cm}=0.98$ fm$^2$~\cite{CiofidegliAtti:1995qe}.

Their relative momentum distribution is expressed as the sum of two Gaussians,
\begin{equation}
n_\text{rel}^{pn}(p)=\frac{C^A}{4\pi}\left[A_1\exp(-\alpha_1 p^2)+A_2\exp(-\alpha_2p^2)\right],
\end{equation}
with the scaling factor $C^A=4.714$~\cite{JeffersonLabHallA:2022ljj}. By requiring that the correlated momentum distribution matches that of \isotope[40][20]{Ca} from Ref.~\cite{CiofidegliAtti:1995qe}, the parameter values are determined to be $A_1=0.23444$ fm$^3$, $\alpha_1=3.2272$ fm$^2$, $A_2=0.006989$ fm$^3$, $\alpha_2=0.23308$ fm$^2$~\cite{JeffersonLabHallA:2022ljj}.

\begin{figure}
\centering
    \includegraphics[width=0.95\columnwidth]{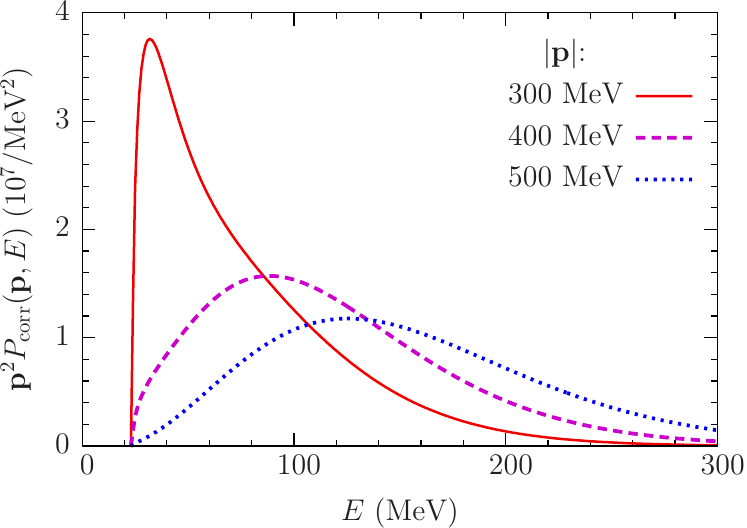}
\caption{\label{fig:SF_fixedP}Dependence of the correlated spectral function of neutrons in argon on removal energy $E$, obtained for fixed values of their momenta $\n p$. For clearer presentation, the results are multiplied by a~factor $\ve p^2$, which also appears in the inclusive cross section calculations.
}
\end{figure}


\begin{figure}
\centering
    \includegraphics[width=0.95\columnwidth]{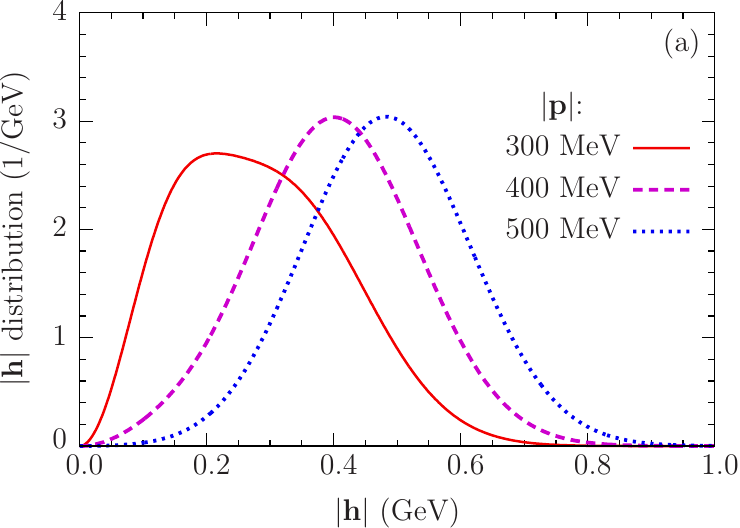}
    \includegraphics[width=0.95\columnwidth]{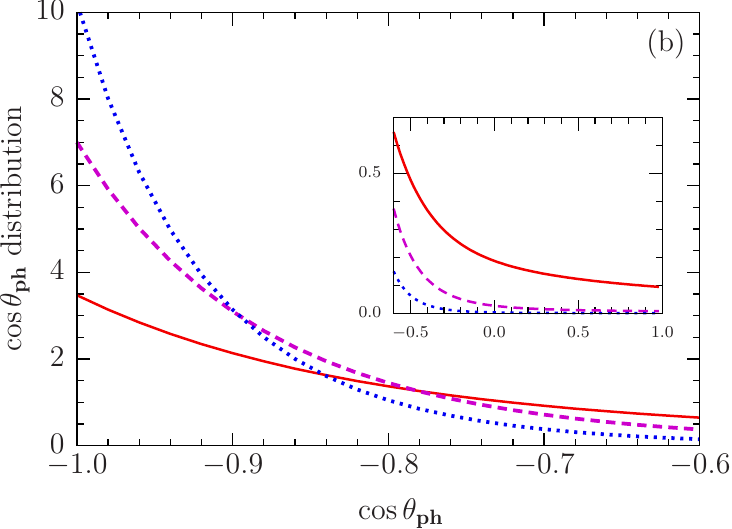}
\caption{\label{fig:correlatedNucleon}Probability distributions for a~proton correlated with the struck neutron of momentum $\n p$, presented as a function of (a) the correlated proton's momentum $\n h$, and (b) the cosine of the opening angle between the two nucleons' momenta.
}
\end{figure}

As the integrand in Eq.~\eqref{eq:corrSF} features cylindrical symmetry about the axis defined by $\ve p$, the azimuthal angle can be readily integrated out, yielding
\begin{equation}\label{eq:corrSF_simplified}
\begin{split}
P_\text{corr}(\ve p,E) &= 2\pi{\mathcal M}\int d\cos\theta_{\ve p\ve h} d\n h\sum_{i=\pm}\delta(\n h - h_i)\\
&\qquad\times \frac{{\n h}^2}{\sqrt\Delta}n^{pn}_\text{cm}(|\ve p +\ve h|)n_\text{rel}^{pn}\left(\Big|\frac{\ve p - \ve h}2\Big|\right),
\end{split}
\end{equation}
where $\theta_{\ve p\ve h}$ denotes the angle between $\ve p$ and $\ve h$, and the correlated nucleon's allowed momenta are
\[
h_\pm =-\frac{\n p}{A-1}\cos\theta_{\ve p\ve h}\pm \sqrt{\Delta},
\]
with
\[
\Delta = \frac{{\n p}^2}{(A-1)^2}(\cos^2\theta_{\ve p\ve h}-1)+2{\mathcal M}(E - E_\text{thr}).
\]

The correlated part of the spectral function of neutrons in argon, obtained from Eq.~\eqref{eq:corrSF_simplified}, is presented in Fig.~\ref{fig:SF_fixedP}. Unlike the MF part, it exhibits a~correlation between the struck nucleon's momentum and removal energy. Additionally, whereas the MF contribution predominantly covers the $\n p \lesssim 300$ MeV and $E \lesssim 80$ MeV region~\cite{JeffersonLabHallA:2022cit,JeffersonLabHallA:2022ljj}, the correlated part receives sizable strength from a~much broader kinematic swath, corresponding to $\n p \leq 1000$ MeV and $E \leq 800$ MeV in our analysis.

The main novel aspect of our implementation of the correlated SF lies in the use of Eq.~\eqref{eq:corrSF_simplified} to also simulate the correlated nucleon's kinematics.

As shown in Fig.~\ref{fig:correlatedNucleon}, while the nucleons in the correlated pair tend to have momenta of similar magnitude and opposing directions, it would be a gross oversimplification to assume that they are in the back-to-back configuration ($\cos\theta_{\ve p\ve h}=-1$) with $\n h =\n p$. Instead, both the relative and the center of mass motions play an important role, and their interplay leads to the obtained distributions.

While in this analysis, our implementation of the description of correlated nucleons is limited to argon, due to its importance for the short-baseline program at Fermilab~\cite{MicroBooNE:2015bmn}, we plan to extend our approach to other targets, such as carbon and oxygen, in the future.

\section{Results and discussion}\label{sec:results}
The large body of existing electron-scattering data for carbon, recently reviewed in Ref.~\cite{Ankowski:2020qbe} and parametrized in Ref.~\cite{Kowal:2023dcq}, offers an excellent opportunity to perform clear-cut tests of the newly implemented \nuwro{} features at the inclusive level~\cite{Ankowski:2022thw}. Because non-QE channels cannot currently be run in the electron mode, we only analyze QE scattering.

\begin{figure*}
\centering
    \subfigure{\label{fig:electrons_a}}
    \subfigure{\label{fig:electrons_b}}
    \subfigure{\label{fig:electrons_c}}
    \includegraphics[width=0.3\textwidth]{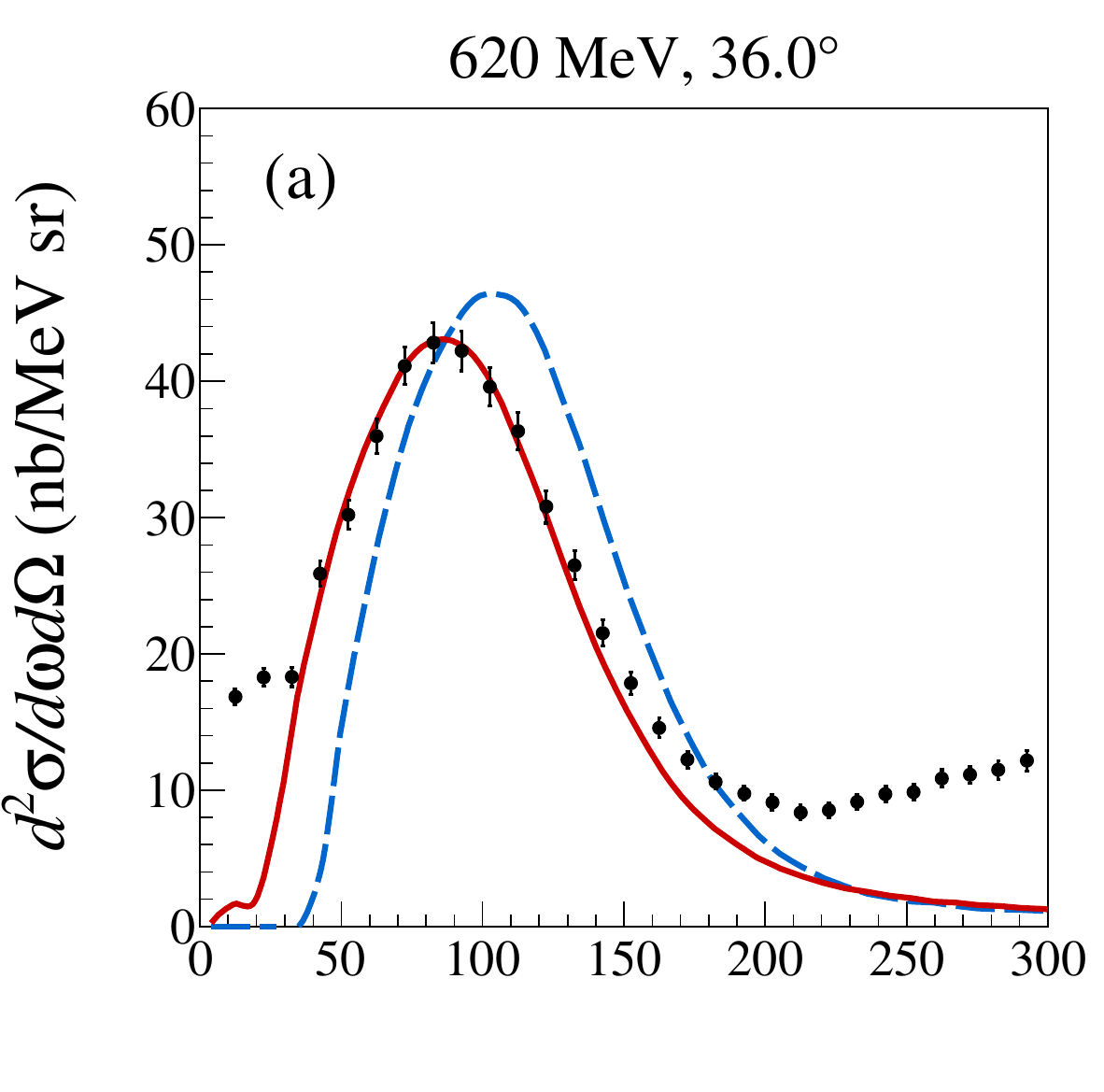}
    \includegraphics[width=0.3\textwidth]{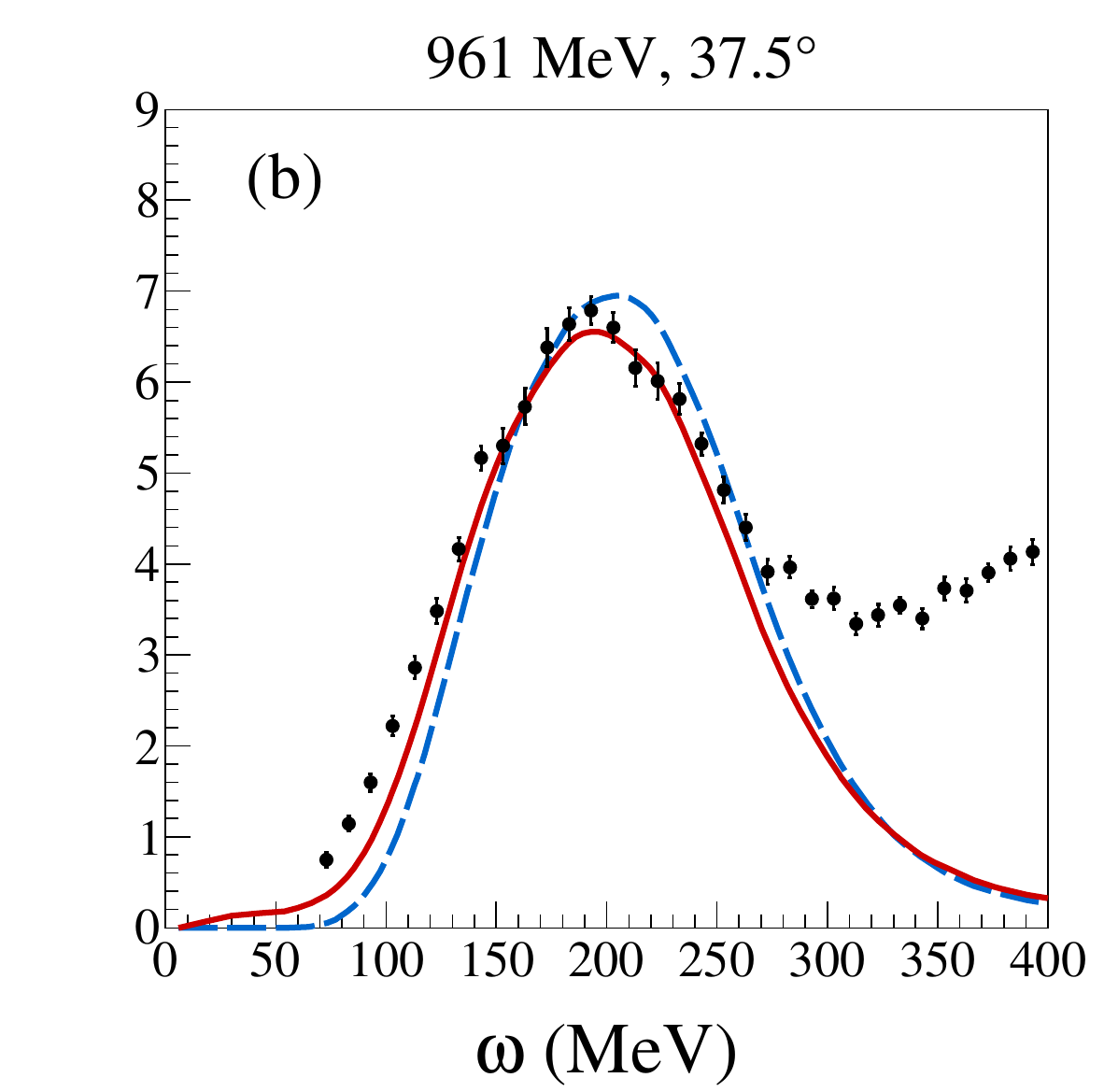}
    \includegraphics[width=0.3\textwidth]{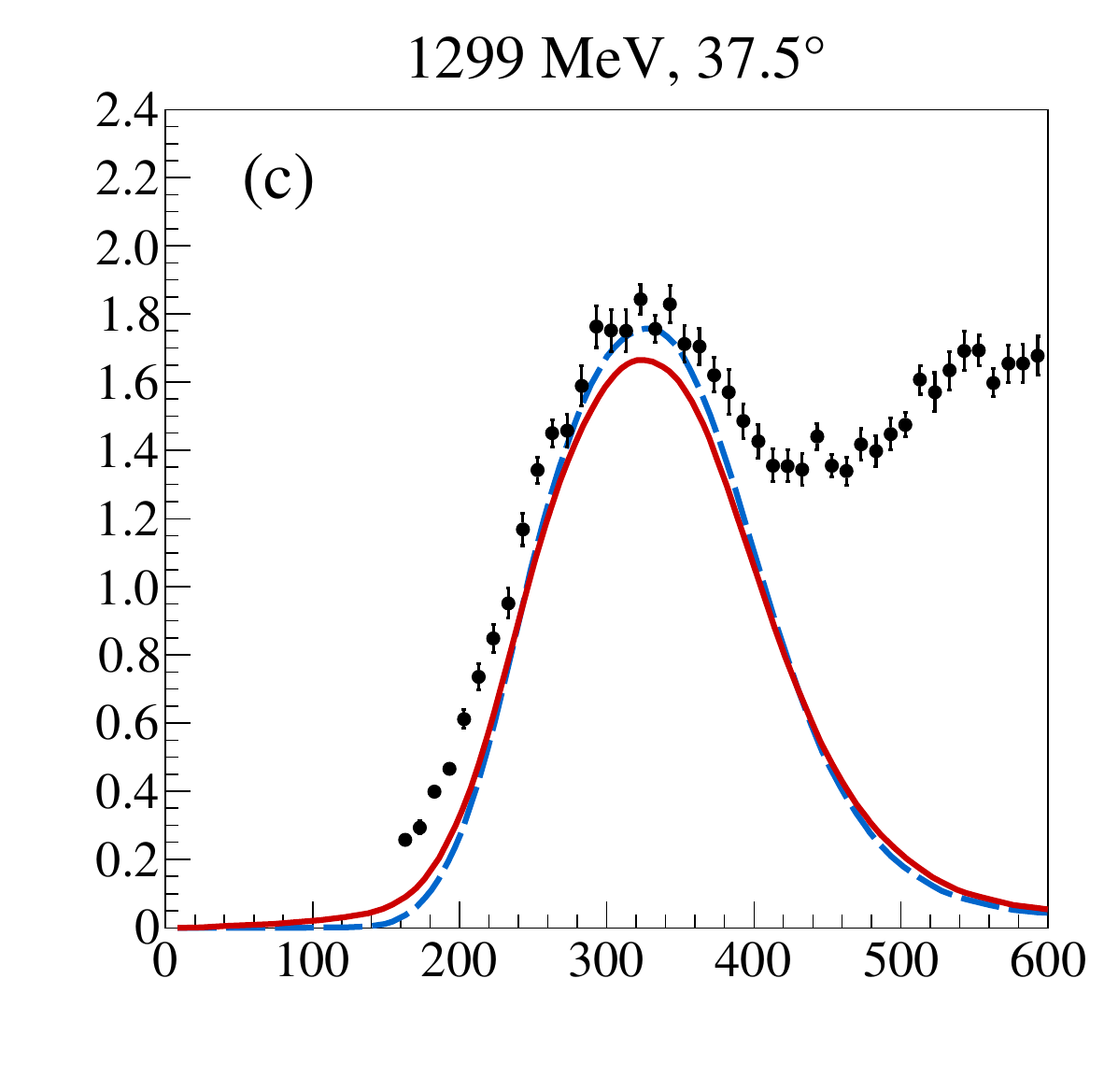}
\caption{\label{fig:electrons}Double differential cross sections for electron scattering on carbon calculated using the SF approach in \nuwro{}. The results obtained with (solid lines) and without (long-dashed lines) FSI are compared to the experimental data reported by (a)~Barreau \etal{}~\cite{Barreau:1983ht}, (b) and (c) Sealock \etal{}~\cite{Sealock:1989nx}. The panels are labeled according to beam energy and scattering angle.
}
\end{figure*}

In Fig.~\ref{fig:electrons}, we compare the \nuwro{} calculations performed with and without FSI.\footnote{To determine the electron-scattering cross sections, we generated $5\times10^8$ events whose cosine of the scattering angle differed by no more than 0.0015 from the desired one.} The kinematic settings of these comparisons are selected in such a way that the scattering angle is---as much as possible---kept fixed, and only the beam energy changes.

Figure~\ref{fig:electrons_a} shows that where the QE and $\Delta$ peaks are clearly separated, including FSI effects allows us to reproduce the experimental data~\cite{Barreau:1983ht} with excellent accuracy. This achievement is only possible thanks to accounting for both the in-medium modification of the struck nucleon spectrum and the redistribution of the cross section, induced by energy exchanges between the struck nucleon and the spectator system. The former effect, described by the real part of the optical potential, produces a~shift of the QE position. The latter redistributes the cross section's strength from its maximum to the tails, resulting in a lower and broader peak. By comparing our result to the data of Fig.~\ref{fig:electrons_a}, we indirectly test the accuracy of the employed $U_V(\tk)=-17$ MeV and $T_A(\tk)=0.76$ values at $\tk=69$ MeV.

At the kinematics of Fig.~\ref{fig:electrons_b}, corresponding to $\tk=168$ MeV, the cross section accounting for FSI still agrees with the data~\cite{Sealock:1989nx} very well. Upon close examination one can, however, see that $U_V(\tk)=-6.3$ MeV seems to be somewhat insufficient, and a~potential deeper by some 7 MeV would be more accurate. The contribution of non-QE interaction mechanisms is much more significant here than in Fig.~\ref{fig:electrons_a}.

When the beam energy increases further, the agreement between our result with FSI and the data~\cite{Sealock:1989nx} remains very good, both in terms of the shape and the strength of the QE peak, as illustrated in Fig.~\ref{fig:electrons_c}. Nevertheless, at $\tk=289$ MeV, the modest shift produced by $U_V(\tk)=-0.3$ MeV is too small by $\sim$15 MeV. At this kinematics, non-QE scattering extends well into the QE peak.

Our findings suggest that nucleon optical potentials determined from elastic scattering data may not be accurate in the presence of inelastic or two-body interaction mechanisms. It is important to note that this issue may only affect interactions corresponding to high kinetic energies, where even a $\sim$15 MeV effect amounts to a small modification of the knocked-out nucleon's final energy. We will return to this observation when discussing uncertainties of our approach in application to neutrino cross sections.

\begin{figure}[b]
\centering
    \includegraphics[width=0.95\columnwidth]{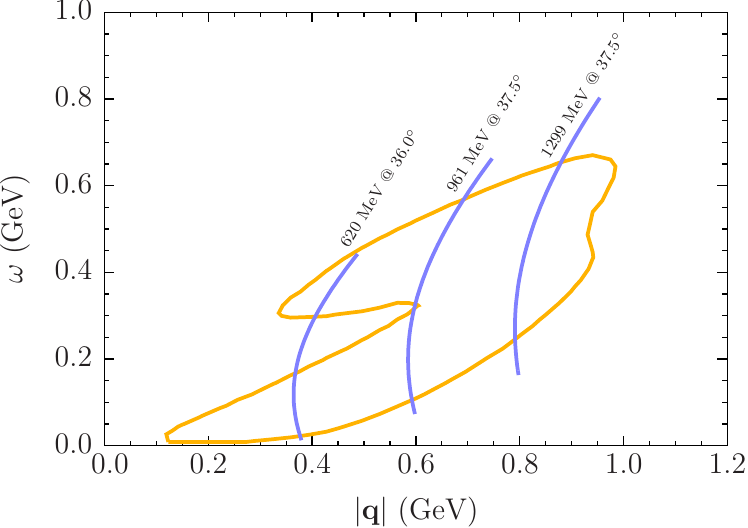}
\caption{\label{fig:contour}Kinematics of the electron-scattering data of Fig.~\ref{fig:electrons}, presented in the plane of momentum transfer and energy transfer, superimposed on the contour representing 68\% of \MB{} CC events, as predicted by \nuwro{}.
}
\end{figure}

The kinematics of our validations are not chosen arbitrarily. In Fig.~\ref{fig:contour} we demonstrate that these datasets uniformly sample the kinematic region containing 68\% of CC events in \MB{}. As a consequence, the observed agreement gives us reasons to expect similarly accurate predictions for quasielastic contributions to neutrino cross sections reported from this experiment.

\begin{figure}
\centering
    \includegraphics[width=0.95\columnwidth]{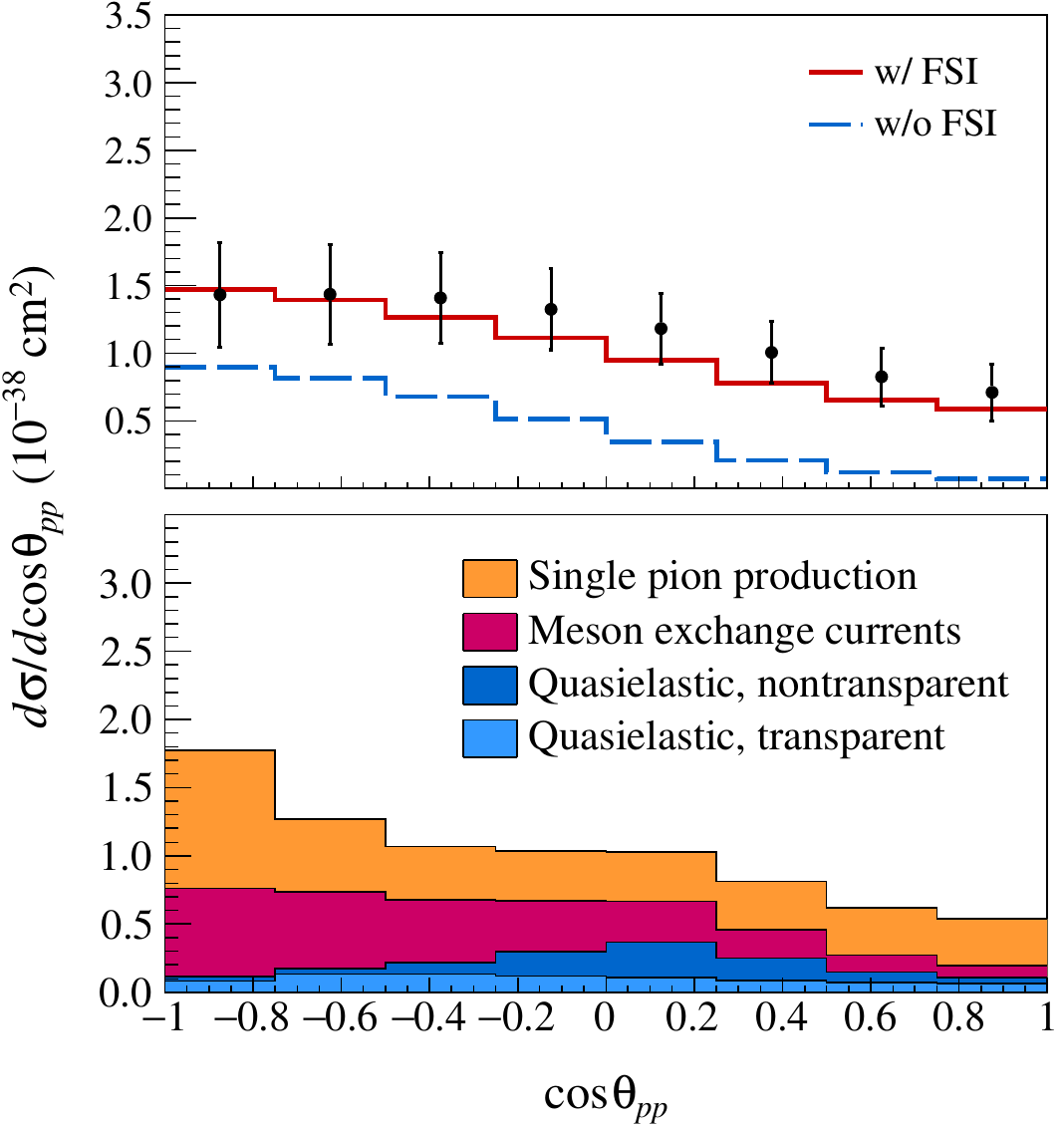}
\caption{\label{fig:cosThetaLR}\MB{} \pp{} differential cross section \dsdcosThetaLR{} calculated within the SF approach in \nuwro{}. Top: The results obtained with final state interactions and correlated protons (``w/ FSI'') and without them (``w/o FSI'') are compared with the experimental data from Ref.~\cite{MicroBooNE:2022emb}. Bottom: Breakdown of the full unsmeared calculations into interaction channels.
}
\end{figure}

Thanks to the \MB{} \pp{} measurement from Ref.~\cite{MicroBooNE:2022emb}, which requires events to contain exactly two protons with true momenta between 0.3 and 1.0 GeV, we can directly test the features newly implemented in \nuwro{}. Figure~\ref{fig:cosThetaLR} compares our calculations performed both with and without FSI and correlated protons to the experimental data for the differential cross section as a~function of \cosThetaLR{}, where \thetaLR{} is the opening angle between the momenta of the two protons. The top panel presents the smeared results, whereas the bottom panel breaks down the unsmeared predictions into channels. Note that in non-QE channels, the sole FSI effect accounted for is the intranuclear cascade.

Our full calculations suggest that the \MB{} data receive substantial contributions from undetected or absorbed pions, MEC, and (mainly nontransparent) quasielastic interactions. As a consequence, the result without FSI and correlated nucleons is unable to reproduce both the normalization and the shape of the experimental dataset, whereas the updated \nuwro{} agrees with it very well. We quantify this observation in Table~\ref{tab:CC2p0pi}.

\begin{table}
\caption{\label{tab:CC2p0pi} $\chi^2$ values per degree of freedom for the agreement between \nuwro{}'s SF simulations and the \MB{} \pp{} data~\cite{MicroBooNE:2022emb}. We compare the results obtained with final state interactions and correlated protons (``w/ FSI'') and without them (``w/o FSI'').}
\begin{tabular*}{\linewidth}{@{\extracolsep{\fill}} l c c c}
    \toprule
        & \dsdcosThetaLR{}  & \dsdcosThetaMuPP{}  & \dsdpT{}  \\
        \midrule
    w/ FSI   & 1.07 \phantom{0}(8.6/8) & 2.54 (20.3/8) & 0.82 \phantom{0}(5.7/7)\\
    w/o FSI  & 2.68 (21.4/8) & 6.21 (49.7/8) & 2.33 (16.3/7)\\
    \bottomrule
\end{tabular*}
\end{table}

\begin{figure}
\centering
    \includegraphics[width=0.95\columnwidth]{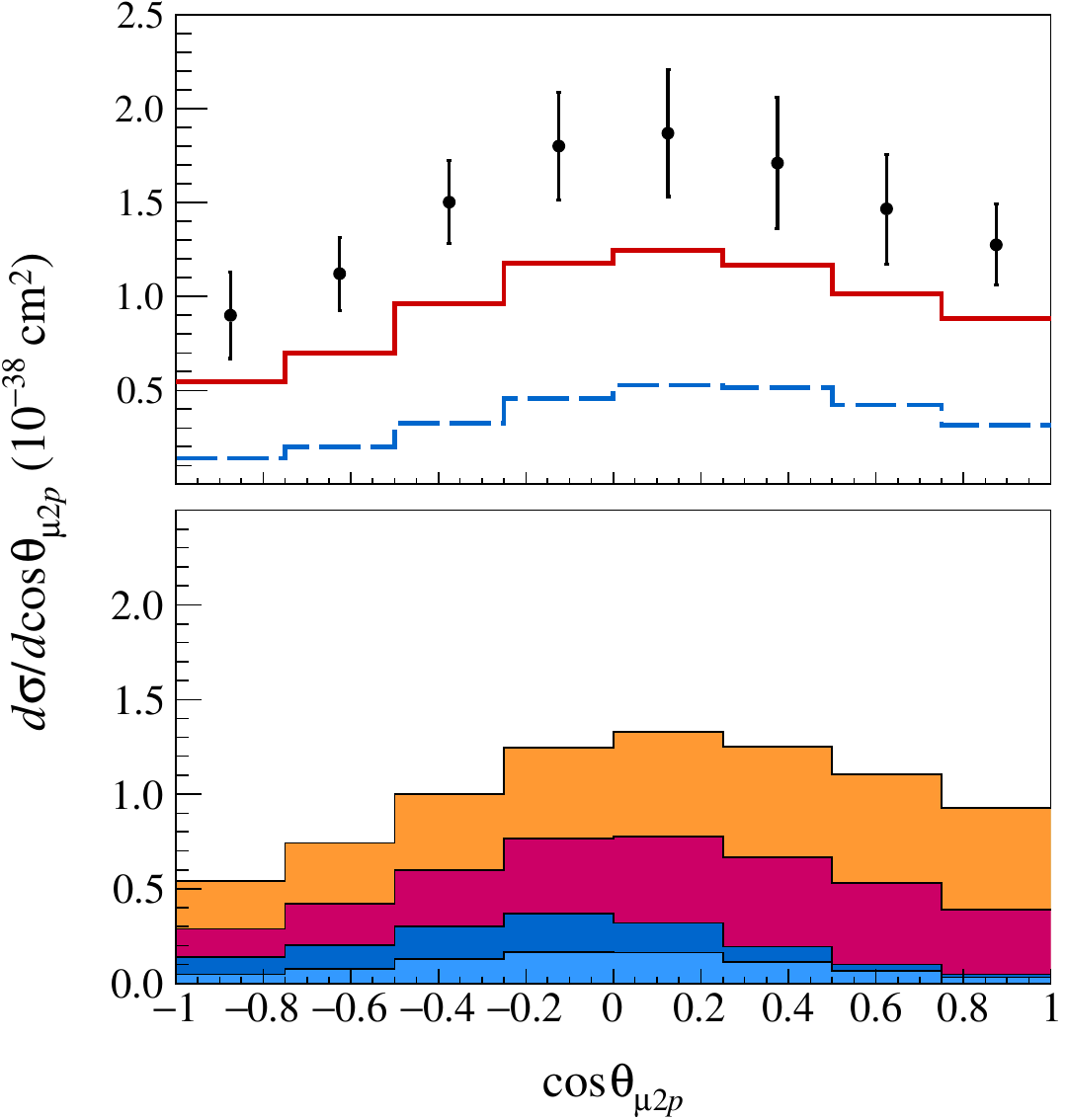}
\caption{\label{fig:cosThetaMuPP}Same as Fig.~\ref{fig:cosThetaLR} but for \dsdcosThetaMuPP{}.
}
\end{figure}

The difference between the results with and without FSI and correlated protons turns out to be even larger for the differential cross section as a function of $\cosThetaMuPP$, where $\thetaMuPP$ denotes the angle between the muon momentum and the total momentum carried by the two protons. Figure~\ref{fig:cosThetaMuPP} shows that while the full calculations are short of reproducing the normalization of the experimental data~\cite{MicroBooNE:2022emb}, they bring important progress toward it. The shape agreement is much better, as reflected by the $\chi^2$ values given in Table~\ref{tab:CC2p0pi}. As a side note, the higher integrated strength of the $d\sigma/d\cosThetaMuPP$ data stems from independent 1D unfolding and variable-specific regularization smearing across phase-space boundaries.

\begin{figure}
\centering
    \includegraphics[width=0.96\columnwidth]{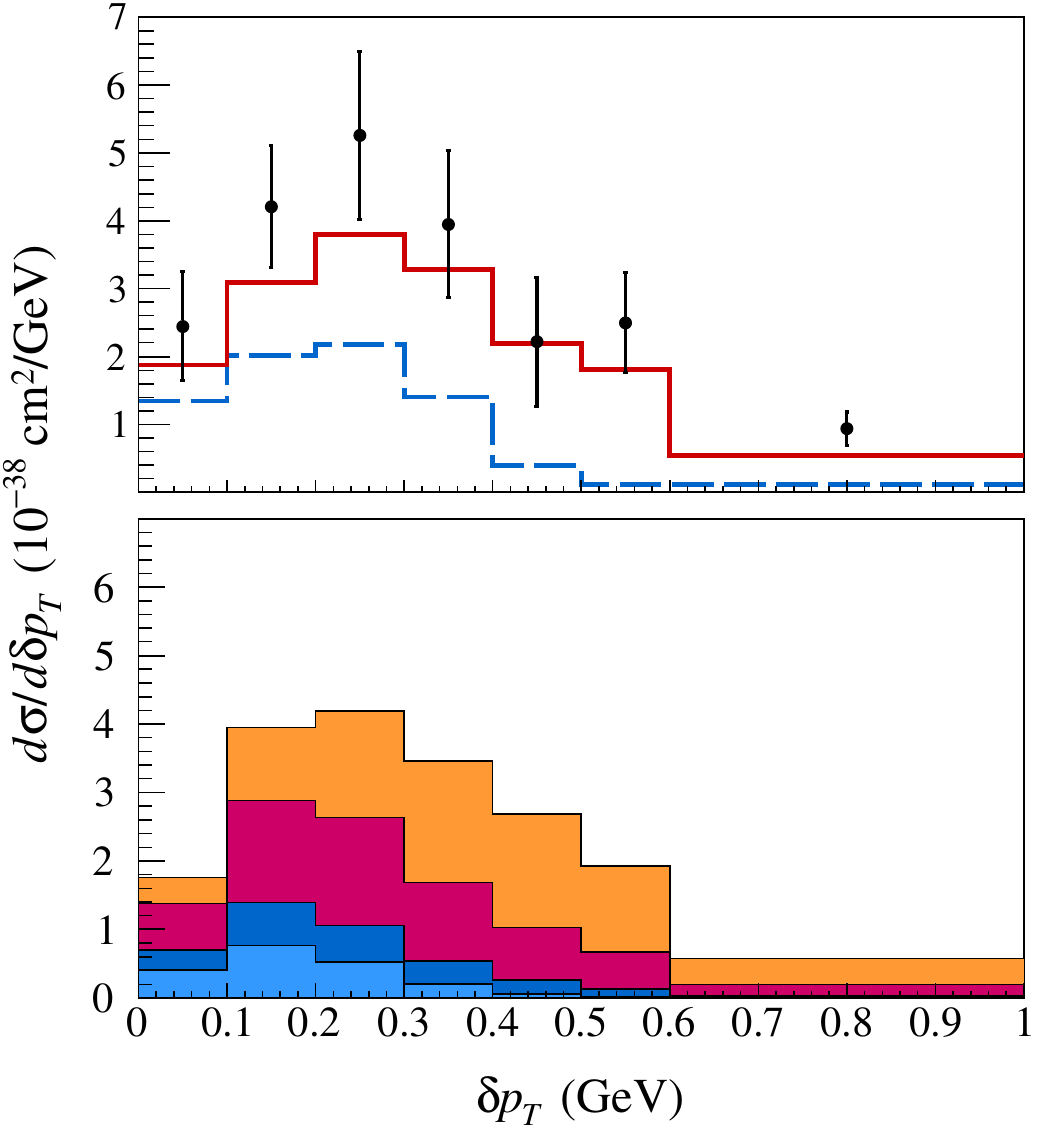}
\caption{\label{fig:P_T}Same as Fig.~\ref{fig:cosThetaLR} but for \dsdpT{}.
}
\end{figure}

We complete the analysis of the \pp{} data by presenting in Fig.~\ref{fig:P_T} the differential cross section as a function of the norm of the transverse momentum imbalance $\dpT=\n{\delta{\bf p}_T}$, where $\delta{\bf p}_T$ is the sum of the muon's and the two protons' momentum components perpendicular to the neutrino direction. While our full calculation seems to somewhat underestimate the cross section's normalization, it provides a very good description of the experimental data, unlike the calculation without FSI and correlated protons, see Table~\ref{tab:CC2p0pi}.

Let us now proceed to the analysis of the \MB{} \p{} measurement reported in Ref.~\cite{MicroBooNE:2023tzj}, in which events are required to contain exactly one proton with momentum between 0.3 and 1.0 GeV. Expectedly, these results are dominated by the contribution of uncorrelated nucleons undergoing QE interactions.

\begin{figure}
\centering
    \includegraphics[width=0.95\columnwidth]{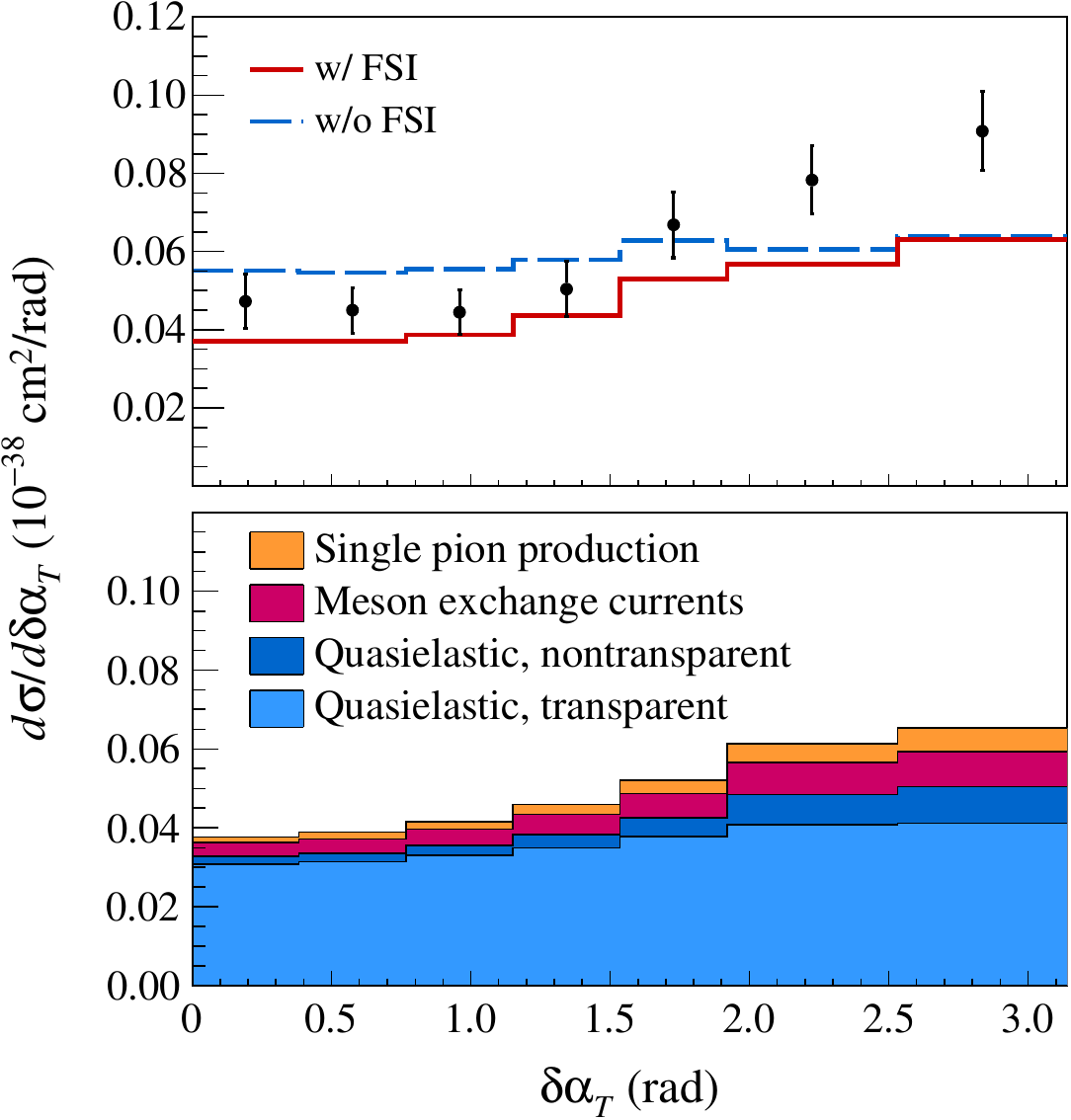}
\caption{\label{fig:dAlphaT}Same as Fig.~\ref{fig:cosThetaLR} but for \MB{} \p{} differential cross section \dsdalpha{}~\cite{MicroBooNE:2023tzj}.
}
\end{figure}

In Fig.~\ref{fig:dAlphaT}, we present the differential cross section as a function of the transverse boosting angle \dalpha{},
\[
\dalpha{}=\arccos\frac{\ve q^T\cdot \delta{\bf p}_T}{| {\bf q}^T| |\delta{\bf p}_T|},
\]
spanned by the transverse momentum transfer $\ve q^T$ and the transverse momentum imbalance $\delta{\bf p}_T$. In this case, $\delta{\bf p}_T$ is defined as the sum of the muon's and proton's momentum components perpendicular to the neutrino direction. Our results show that in the SF approach without FSI and correlated protons, this distribution is rather flat, as observed for the local Fermi gas model in Ref.~\cite{Lu:2015tcr}.

\begin{table}
\caption{\label{tab:CC1p0pi} Same as in Table~\ref{tab:CC2p0pi} but for the \MB{} \p{} data~\cite{MicroBooNE:2023tzj}.}
\begin{tabular*}{\linewidth}{@{\extracolsep{\fill}} l c c c}
    \toprule
        & \dsdalpha{}  & \dsdpT{}  & \dsdpTx{}  \\
        \midrule
    w/ FSI   & \phantom{0}1.58 (11.1/7) & 1.65 (21.5/13) & 1.31 (14.5/11) \\
    w/o FSI  & \phantom{0}7.32 (51.2/7) & 5.02 (65.3/13) & 2.86 (31.5/11) \\
    \bottomrule
\end{tabular*}
\end{table}

On the other hand, the shape of the experimental data exhibits a clear $\dalpha{}$ dependence, much better reproduced by our full calculations. This is possible thanks to the correlated SF, MEC, and pion contributions, all of which contribute to the upturn at high \dalpha{}, correlated with high \dpT{}. We observe that the remaining deficit of the cross section occurs largely in the high \dalpha{} and high \dpT{} region, see Supplemental Material~\cite{SupplemementalMaterial}. The kinematics corresponding to the deficit indicates that FSI are still underestimated in our calculations. We will return to this issue when discussing the uncertainties of our approach. Nevertheless, the experimental $\dsdalpha{}$ cross section is reproduced rather well, as quantified in Table~\ref{tab:CC1p0pi}.

\begin{figure}
\centering
    \includegraphics[width=0.96\columnwidth]{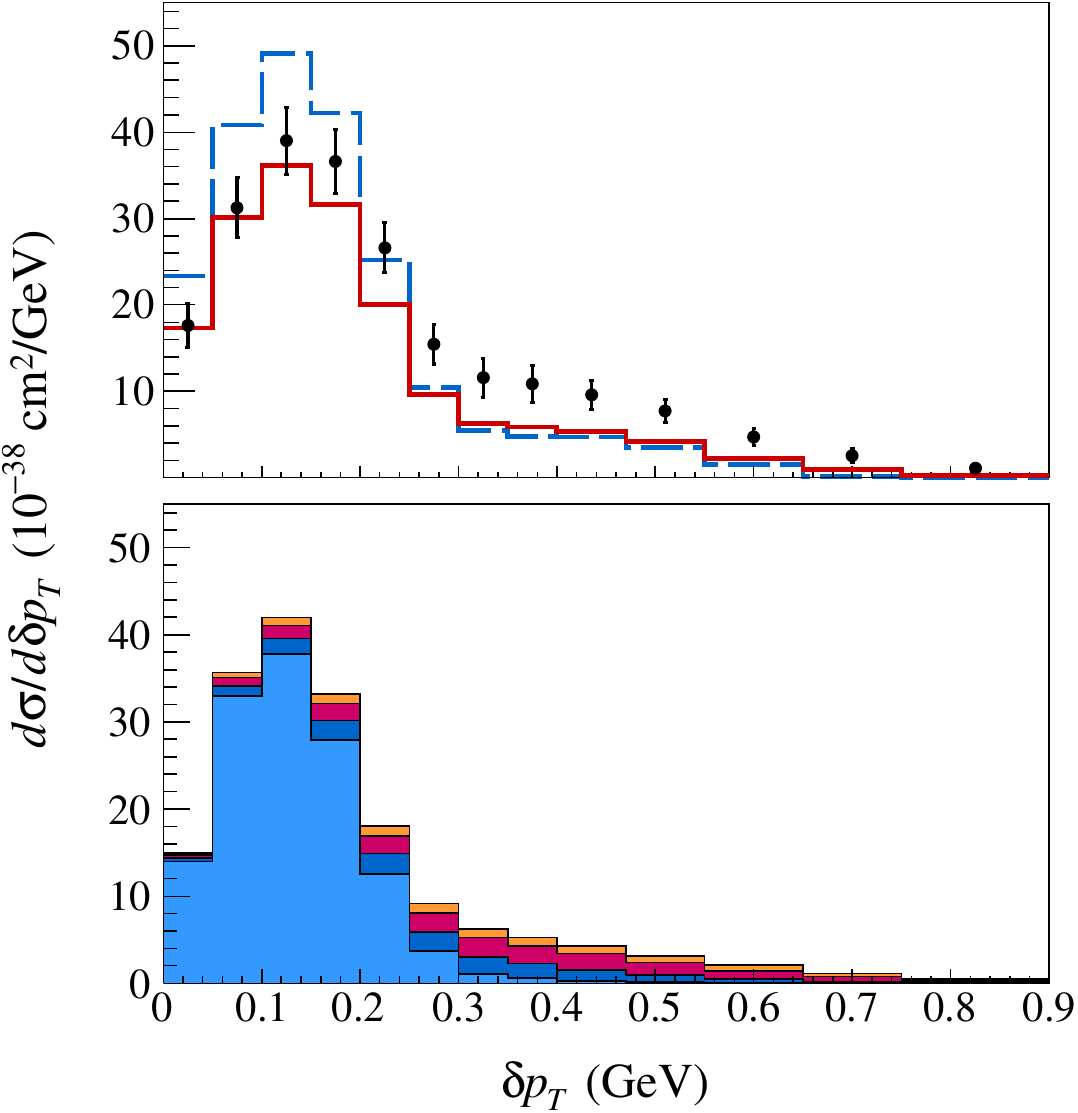}
\caption{\label{fig:dPT}Same as Fig.~\ref{fig:cosThetaLR} but for \MB{} \p{} differential cross section \dsdpT{}~\cite{MicroBooNE:2023tzj}.
}
\end{figure}

Complementary information about \p{} interactions is provided by the \dsdpT{} results, presented in Fig.~\ref{fig:dPT}. The most apparent effect of including FSI and correlated protons is the reduction of the peak centered at $\dpT{}\sim125$ MeV, supported by the experimental data~\cite{MicroBooNE:2023tzj}. In this region, dominated by the MF contribution, FSI can suppress the \p{} cross section by producing additional protons with momentum above 0.3 GeV. In the tail, dominated by the correlated SF in the QE channel and by MEC, FSI plays a twofold role: absorption of correlated protons may increase the cross section, whereas FSI of the struck nucleon may reduce it, by contributing additional protons that do not satisfy the \p{} selection criteria.

The deficit of the cross section in the tail suggests that the absorptive processes may be too weak in \nuwro{}'s intranuclear cascade. In spite of this issue, our full calculation reproduces the experimental data reasonably well, see Table~\ref{tab:CC1p0pi}. For completeness, we also provide the result for the \dsdpTx{} cross section.

Our treatment of FSI in the argon nucleus requires knowledge of its nuclear transparency and the real part of its optical potential. As these quantities are subject to uncertainties, a natural question arises: how significant are these uncertainties for the analyzed neutrino cross sections?

We previously estimated the nuclear transparency to be known to $\pm3\%$. This uncertainty has a minor to moderate effect on the considered cross sections, as quantified in Table~\ref{tab:unc}. We typically observe a~decrease (increase) in $\chi^2$ when the transparency is lowered (increased), consistent with our previous conclusion that absorption may be underestimated in \nuwro{}'s intranuclear cascade.

To gauge the sensitivity of our results to the uncertainty of the real part of the optical potential, we cap its maximal value,
\begin{equation}\label{eq:UMod}
U_V'(\tk)=\min\{U_V(\tk),-15 \text{ MeV}\}.
\end{equation}

\begin{table}
\caption{\label{tab:unc} Sensitivity of the goodness of fit to variations in FSI model parameters. We present the impact on $\chi^2$ per degree of freedom when the nuclear transparency $T_A$ is varied by $\pm3\%$ and the real part of the optical potential is modified according to Eq.~\eqref{eq:UMod}.}
\begin{tabular*}{\linewidth}{@{\extracolsep{\fill}} l c c c c}
    \toprule
        & Default & $0.97\times T_A$ & $1.03\times T_A$  & $U_V'$  \\
    \midrule
    \pp{}: & & & &\\
    \dsdcosThetaLR{}   & 1.07 & 0.86 & 1.08 & 1.03\\
    \dsdcosThetaMuPP{} & 2.54 & 2.40 & 2.75 & 2.51\\
    \dsdpT{}           & 0.82 & 0.84 & 0.83 & 0.85\\[3pt]
    \p{}: & & & &\\
    \dsdalpha{}      & 1.58  & 1.55 & 1.63 & 0.97\\
    \dsdpT{}         & 1.65  & 1.62 & 1.79 & 1.55\\
    \dsdpTx{}        & 1.31  & 1.30 & 1.36 & 1.44\\
    \bottomrule
\end{tabular*}
\end{table}


This modification is a pragmatic attempt to account for our observation that $U_V(\tk)$ becomes too shallow for high $\tk$ values, where inelastic and two-body interaction mechanisms begin to contribute in the region of the QE peak. Being based on comparisons with the carbon data presented in Fig.~\ref{fig:electrons}---in the absence of the broad body of electron-scattering data for argon that our community urgently needs---the potential~\eqref{eq:UMod} should be viewed as a desperate remedy for desperate times.

It turns out that the \dsdalpha{} cross section is the most sensitive to the difference between the $U_V$ and $U_V'$ potentials, and that using $U_V'$ lowers $\chi^2$ per degree of freedom to $\sim$1. Other considered cross sections exhibit weak to moderate sensitivity to the change in the real part of the optical potential, as shown in Table~\ref{tab:unc}.

\section{Summary}\label{sec:summary}
To address the needs of accelerator-based neutrino-oscillation experiments, we implement a consistent treatment of multinucleon final states in quasielastic scattering at the inclusive and exclusive level in \nuwro{}'s SF approach. For targets ranging from carbon to iron, we implement a description of FSI by employing the convolution scheme. For argon, we also model additional protons resulting from short-range correlations in the initial nuclear state, having in mind the applications in the short-baseline program at Fermilab.

We validate the accuracy of our FSI approach in the quasielastic channel against inclusive electron-scattering data for carbon, covering \MB{} kinematics. For neutrino interactions, we test our simulations by comparing them with the \pp{} and \p{} cross sections reported by the \MBC{}. We find that accounting for FSI and correlated nucleons is essential to bring the \nuwro{} predictions into good agreement with the experimental data, and that the contributions of non-QE channels are sizable. Finally, we analyze the sensitivity of our FSI description to the uncertainties in its ingredients: the nuclear transparency and the real part of the optical potential.

Our results suggest that \nuwro{}'s intranuclear cascade may be currently underestimating the absorptive processes. Additionally, we find that nucleon optical potentials determined from elastic scattering data may not be accurate in the presence of other interaction mechanisms. To address this problem, our community urgently needs new measurements of the cross sections for inclusive electron scattering---particularly on oxygen and argon---covering the kinematics relevant to the accelerator-based neutrino-oscillation experiments. Achieving this goal is a~prerequisite for realizing the full physics potential of next-generation neutrino-oscillation programs.



\begin{acknowledgments}
We are deeply indebted to Daniel Barrow, London Cooper-Troendle, Nitish Nayak, and Afroditi Papadopoulou for sharing the details of the \MB{} results. We would also like to thank Shinichi Hama for providing us with the parametrization of optical potentials in the \textsc{global} code. This work is partly (A.M.A., K.M.G., and J.T.S.) or fully (R.D.B., J.L.B., B.E.K., and H.P.) supported by the National Science Centre under grant UMO-2021/41/B/ST2/02778.
\end{acknowledgments}
\newpage





\begin{thebibliography}{50}%
\makeatletter
\providecommand \@ifxundefined [1]{%
 \@ifx{#1\undefined}
}%
\providecommand \@ifnum [1]{%
 \ifnum #1\expandafter \@firstoftwo
 \else \expandafter \@secondoftwo
 \fi
}%
\providecommand \@ifx [1]{%
 \ifx #1\expandafter \@firstoftwo
 \else \expandafter \@secondoftwo
 \fi
}%
\providecommand \natexlab [1]{#1}%
\providecommand \enquote  [1]{``#1''}%
\providecommand \bibnamefont  [1]{#1}%
\providecommand \bibfnamefont [1]{#1}%
\providecommand \citenamefont [1]{#1}%
\providecommand \href@noop [0]{\@secondoftwo}%
\providecommand \href [0]{\begingroup \@sanitize@url \@href}%
\providecommand \@href[1]{\@@startlink{#1}\@@href}%
\providecommand \@@href[1]{\endgroup#1\@@endlink}%
\providecommand \@sanitize@url [0]{\catcode `\\12\catcode `\$12\catcode
  `\&12\catcode `\#12\catcode `\^12\catcode `\_12\catcode `\%12\relax}%
\providecommand \@@startlink[1]{}%
\providecommand \@@endlink[0]{}%
\providecommand \url  [0]{\begingroup\@sanitize@url \@url }%
\providecommand \@url [1]{\endgroup\@href {#1}{\urlprefix }}%
\providecommand \urlprefix  [0]{URL }%
\providecommand \Eprint [0]{\href }%
\providecommand \doibase [0]{https://doi.org/}%
\providecommand \selectlanguage [0]{\@gobble}%
\providecommand \bibinfo  [0]{\@secondoftwo}%
\providecommand \bibfield  [0]{\@secondoftwo}%
\providecommand \translation [1]{[#1]}%
\providecommand \BibitemOpen [0]{}%
\providecommand \bibitemStop [0]{}%
\providecommand \bibitemNoStop [0]{.\EOS\space}%
\providecommand \EOS [0]{\spacefactor3000\relax}%
\providecommand \BibitemShut  [1]{\csname bibitem#1\endcsname}%
\let\auto@bib@innerbib\@empty
\bibitem [{\citenamefont {Abe}\ \emph {et~al.}(2015)\citenamefont {Abe} \emph
  {et~al.}}]{Hyper-KamiokandeProto-:2015xww}%
  \BibitemOpen
  \bibfield  {author} {\bibinfo {author} {\bibfnamefont {K.}~\bibnamefont
  {Abe}} \emph {et~al.} (\bibinfo {collaboration} {Hyper-Kamiokande
  Proto-Collaboration}),\ }\href {https://doi.org/10.1093/ptep/ptv061}
  {\bibfield  {journal} {\bibinfo  {journal} {Prog. Theor. Exp. Phys.}\
  }\textbf {\bibinfo {volume} {2015}},\ \bibinfo {pages} {053C02} (\bibinfo
  {year} {2015})}\BibitemShut {NoStop}%
\bibitem [{\citenamefont {Abi}\ \emph {et~al.}(2020)\citenamefont {Abi} \emph
  {et~al.}}]{DUNE:2020lwj}%
  \BibitemOpen
  \bibfield  {author} {\bibinfo {author} {\bibfnamefont {B.}~\bibnamefont
  {Abi}} \emph {et~al.} (\bibinfo {collaboration} {DUNE Collaboration}),\
  }\href {https://doi.org/10.1088/1748-0221/15/08/T08008} {\bibfield  {journal}
  {\bibinfo  {journal} {J. Instrum.}\ }\textbf {\bibinfo {volume} {15}},\
  \bibinfo {pages} {T08008} (\bibinfo {year} {2020})}\BibitemShut {NoStop}%
\bibitem [{\citenamefont {Acero}\ \emph {et~al.}(2022)\citenamefont {Acero}
  \emph {et~al.}}]{NOvA:2021nfi}%
  \BibitemOpen
  \bibfield  {author} {\bibinfo {author} {\bibfnamefont {M.~A.}\ \bibnamefont
  {Acero}} \emph {et~al.} (\bibinfo {collaboration} {NOvA Collaboration}),\
  }\href {https://doi.org/10.1103/PhysRevD.106.032004} {\bibfield  {journal}
  {\bibinfo  {journal} {Phys. Rev. D}\ }\textbf {\bibinfo {volume} {106}},\
  \bibinfo {pages} {032004} (\bibinfo {year} {2022})}\BibitemShut {NoStop}%
\bibitem [{\citenamefont {Abe}\ \emph {et~al.}(2023)\citenamefont {Abe} \emph
  {et~al.}}]{T2K:2023smv}%
  \BibitemOpen
  \bibfield  {author} {\bibinfo {author} {\bibfnamefont {K.}~\bibnamefont
  {Abe}} \emph {et~al.} (\bibinfo {collaboration} {T2K Collaboration}),\ }\href
  {https://doi.org/10.1140/epjc/s10052-023-11819-x} {\bibfield  {journal}
  {\bibinfo  {journal} {Eur. Phys. J. C}\ }\textbf {\bibinfo {volume} {83}},\
  \bibinfo {pages} {782} (\bibinfo {year} {2023})}\BibitemShut {NoStop}%
\bibitem [{\citenamefont {Acciarri}\ \emph {et~al.}(2025)\citenamefont
  {Acciarri} \emph {et~al.}}]{SBND:2025lha}%
  \BibitemOpen
  \bibfield  {author} {\bibinfo {author} {\bibfnamefont {R.}~\bibnamefont
  {Acciarri}} \emph {et~al.} (\bibinfo {collaboration} {SBND Collaboration}),\
  }\Eprint {https://arxiv.org/abs/2504.00245} {arXiv:2504. 00245}
  (\bibinfo {year} {2025})\BibitemShut {NoStop}%
\bibitem [{\citenamefont {Acciarri}\ \emph {et~al.}()\citenamefont {Acciarri}
  \emph {et~al.}}]{MicroBooNE:2015bmn}%
  \BibitemOpen
  \bibfield  {author} {\bibinfo {author} {\bibfnamefont {R.}~\bibnamefont
  {Acciarri}} \emph {et~al.} (\bibinfo {collaboration} {MicroBooNE, LAr1-ND,
  ICARUS-WA104 Collaborations}),\ }\Eprint {https://arxiv.org/abs/1503.01520}
  {arXiv:1503.01520} (\bibinfo {year} {2015})\BibitemShut {NoStop}%
\bibitem [{\citenamefont {Abratenko}\ \emph {et~al.}(2026)\citenamefont
  {Abratenko} \emph {et~al.}}]{MicroBooNE:2022emb}%
  \BibitemOpen
  \bibfield  {author} {\bibinfo {author} {\bibfnamefont {P.}~\bibnamefont
  {Abratenko}} \emph {et~al.} (\bibinfo {collaboration} {MicroBooNE}),\ }\href
  {https://doi.org/10.1016/j.physletb.2025.140052} {\bibfield  {journal}
  {\bibinfo  {journal} {Phys. Lett. B}\ }\textbf {\bibinfo {volume} {872}},\
  \bibinfo {pages} {140052} (\bibinfo {year} {2026})}\BibitemShut {NoStop}%
\bibitem [{\citenamefont {Abratenko}\ \emph {et~al.}(2023)\citenamefont
  {Abratenko} \emph {et~al.}}]{MicroBooNE:2023tzj}%
  \BibitemOpen
  \bibfield  {author} {\bibinfo {author} {\bibfnamefont {P.}~\bibnamefont
  {Abratenko}} \emph {et~al.} (\bibinfo {collaboration} {MicroBooNE
  Collaboration}),\ }\href {https://doi.org/10.1103/PhysRevLett.131.101802}
  {\bibfield  {journal} {\bibinfo  {journal} {Phys. Rev. Lett.}\ }\textbf
  {\bibinfo {volume} {131}},\ \bibinfo {pages} {101802} (\bibinfo {year}
  {2023})}\BibitemShut {NoStop}%
\bibitem [{\citenamefont {Nikolakopoulos}\ \emph {et~al.}(2024)\citenamefont
  {Nikolakopoulos}, \citenamefont {Ershova}, \citenamefont
  {Gonz{\'a}lez-Jim{\'e}nez}, \citenamefont {Isaacson}, \citenamefont {Kelly},
  \citenamefont {Niewczas}, \citenamefont {Rocco},\ and\ \citenamefont
  {S{\'a}nchez}}]{Nikolakopoulos:2024mjj}%
  \BibitemOpen
  \bibfield  {author} {\bibinfo {author} {\bibfnamefont {A.}~\bibnamefont
  {Nikolakopoulos}}, \bibinfo {author} {\bibfnamefont {A.}~\bibnamefont
  {Ershova}}, \bibinfo {author} {\bibfnamefont {R.}~\bibnamefont
  {Gonz{\'a}lez-Jim{\'e}nez}}, \bibinfo {author} {\bibfnamefont
  {J.}~\bibnamefont {Isaacson}}, \bibinfo {author} {\bibfnamefont {A.~M.}\
  \bibnamefont {Kelly}}, \bibinfo {author} {\bibfnamefont {K.}~\bibnamefont
  {Niewczas}}, \bibinfo {author} {\bibfnamefont {N.}~\bibnamefont {Rocco}},\
  and\ \bibinfo {author} {\bibfnamefont {F.}~\bibnamefont {S{\'a}nchez}},\
  }\href {https://doi.org/10.1103/PhysRevC.110.054611} {\bibfield  {journal}
  {\bibinfo  {journal} {Phys. Rev. C}\ }\textbf {\bibinfo {volume} {110}},\
  \bibinfo {pages} {054611} (\bibinfo {year} {2024})}\BibitemShut {NoStop}%
\bibitem [{\citenamefont {McKean}\ \emph {et~al.}(2025)\citenamefont {McKean},
  \citenamefont {Gonz{\'a}lez-Jim{\'e}nez}, \citenamefont {Kabirnezhad},
  \citenamefont {Ud{\'\i}as},\ and\ \citenamefont {Uchida}}]{McKean:2025khb}%
  \BibitemOpen
  \bibfield  {author} {\bibinfo {author} {\bibfnamefont {J.}~\bibnamefont
  {McKean}}, \bibinfo {author} {\bibfnamefont {R.}~\bibnamefont
  {Gonz{\'a}lez-Jim{\'e}nez}}, \bibinfo {author} {\bibfnamefont
  {M.}~\bibnamefont {Kabirnezhad}}, \bibinfo {author} {\bibfnamefont {J.~M.}\
  \bibnamefont {Ud{\'\i}as}},\ and\ \bibinfo {author} {\bibfnamefont
  {Y.}~\bibnamefont {Uchida}},\ }\href {https://doi.org/10.1103/f7x5-snmz}
  {\bibfield  {journal} {\bibinfo  {journal} {Phys. Rev. D}\ }\textbf {\bibinfo
  {volume} {112}},\ \bibinfo {pages} {032009} (\bibinfo {year}
  {2025})}\BibitemShut {NoStop}%
\bibitem [{\citenamefont {Filali}\ \emph {et~al.}(2025)\citenamefont {Filali},
  \citenamefont {Munteanu},\ and\ \citenamefont {Dolan}}]{Filali:2024vpy}%
  \BibitemOpen
  \bibfield  {author} {\bibinfo {author} {\bibfnamefont {W.}~\bibnamefont
  {Filali}}, \bibinfo {author} {\bibfnamefont {L.}~\bibnamefont {Munteanu}},\
  and\ \bibinfo {author} {\bibfnamefont {S.}~\bibnamefont {Dolan}},\ }\href
  {https://doi.org/10.1103/PhysRevD.111.032009} {\bibfield  {journal} {\bibinfo
   {journal} {Phys. Rev. D}\ }\textbf {\bibinfo {volume} {111}},\ \bibinfo
  {pages} {032009} (\bibinfo {year} {2025})}\BibitemShut {NoStop}%
\bibitem [{\citenamefont {Dolan}\ \emph {et~al.}(2026)\citenamefont {Dolan},
  \citenamefont {Pickering}, \citenamefont {Stowell}, \citenamefont
  {Wilkinson},\ and\ \citenamefont {Wret}}]{Dolan:2026nlr}%
  \BibitemOpen
  \bibfield  {author} {\bibinfo {author} {\bibfnamefont {S.}~\bibnamefont
  {Dolan}}, \bibinfo {author} {\bibfnamefont {L.}~\bibnamefont {Pickering}},
  \bibinfo {author} {\bibfnamefont {P.}~\bibnamefont {Stowell}}, \bibinfo
  {author} {\bibfnamefont {C.}~\bibnamefont {Wilkinson}},\ and\ \bibinfo
  {author} {\bibfnamefont {C.}~\bibnamefont {Wret}},\ }\href@noop {} {\Eprint {https://arxiv.org/abs/2605.28671}
  {arXiv:2605.28671}\ (\bibinfo {year} {2026})}\BibitemShut {NoStop}%
\bibitem [{\citenamefont {Yan}\ \emph {et~al.}(2025)\citenamefont {Yan},
  \citenamefont {Wen}, \citenamefont {Gallmeister}, \citenamefont {Lu},
  \citenamefont {Mosel},\ and\ \citenamefont {Zheng}}]{Yan:2025aau}%
  \BibitemOpen
  \bibfield  {author} {\bibinfo {author} {\bibfnamefont {Q.}~\bibnamefont
  {Yan}}, \bibinfo {author} {\bibfnamefont {K.}~\bibnamefont {Wen}}, \bibinfo
  {author} {\bibfnamefont {K.}~\bibnamefont {Gallmeister}}, \bibinfo {author}
  {\bibfnamefont {X.}~\bibnamefont {Lu}}, \bibinfo {author} {\bibfnamefont
  {U.}~\bibnamefont {Mosel}},\ and\ \bibinfo {author} {\bibfnamefont
  {Y.}~\bibnamefont {Zheng}},\ }\href {https://doi.org/10.1103/yzqx-7ttd}
  {\bibfield  {journal} {\bibinfo  {journal} {Phys. Rev. D}\ }\textbf {\bibinfo
  {volume} {112}},\ \bibinfo {pages} {093007} (\bibinfo {year}
  {2025})}\BibitemShut {NoStop}%
\bibitem [{\citenamefont {Isaacson}\ \emph {et~al.}(2026)\citenamefont
  {Isaacson}, \citenamefont {Jay}, \citenamefont {Lovato}, \citenamefont
  {Machado}, \citenamefont {Nikolakopoulos}, \citenamefont {Rocco},\ and\
  \citenamefont {Steinberg}}]{Isaacson:2025cnk}%
  \BibitemOpen
  \bibfield  {author} {\bibinfo {author} {\bibfnamefont {J.}~\bibnamefont
  {Isaacson}}, \bibinfo {author} {\bibfnamefont {W.}~\bibnamefont {Jay}},
  \bibinfo {author} {\bibfnamefont {A.}~\bibnamefont {Lovato}}, \bibinfo
  {author} {\bibfnamefont {P.}~\bibnamefont {Machado}}, \bibinfo {author}
  {\bibfnamefont {A.}~\bibnamefont {Nikolakopoulos}}, \bibinfo {author}
  {\bibfnamefont {N.}~\bibnamefont {Rocco}},\ and\ \bibinfo {author}
  {\bibfnamefont {N.}~\bibnamefont {Steinberg}},\ }\href
  {https://doi.org/10.1103/13bh-22lm} {\bibfield  {journal} {\bibinfo
  {journal} {Phys. Rev. D}\ }\textbf {\bibinfo {volume} {113}},\ \bibinfo
  {pages} {036005} (\bibinfo {year} {2026})}\BibitemShut {NoStop}%
\bibitem [{\citenamefont {Liu}\ \emph {et~al.}(2026)\citenamefont {Liu},
  \citenamefont {Gardiner},\ and\ \citenamefont {Dytman}}]{Liu:2026wlw}%
  \BibitemOpen
  \bibfield  {author} {\bibinfo {author} {\bibfnamefont {L.}~\bibnamefont
  {Liu}}, \bibinfo {author} {\bibfnamefont {S.}~\bibnamefont {Gardiner}},\ and\
  \bibinfo {author} {\bibfnamefont {S.}~\bibnamefont {Dytman}},\ }\href
  {https://doi.org/10.1016/j.physletb.2026.140750} {\bibfield  {journal}
  {\bibinfo  {journal} {Phys. Lett. B}\ }\textbf {\bibinfo {volume} {880}},\
  \bibinfo {pages} {140750} (\bibinfo {year} {2026})}\BibitemShut
  {NoStop}%
\bibitem [{NuW()}]{NuWro}%
  \BibitemOpen
  \href@noop {} {}\bibinfo {howpublished} {\nuwro{} official repository,
  \url{https://github.com/NuWro/nuwro}}\BibitemShut {NoStop}%
\bibitem [{\citenamefont {Dharmapal~Banerjee}\ \emph
  {et~al.}(2024)\citenamefont {Dharmapal~Banerjee}, \citenamefont {Ankowski},
  \citenamefont {Graczyk}, \citenamefont {Kowal}, \citenamefont {Prasad},\ and\
  \citenamefont {Sobczyk}}]{Banerjee:2023hub}%
  \BibitemOpen
  \bibfield  {author} {\bibinfo {author} {\bibfnamefont {R.}~\bibnamefont
  {Dharmapal~Banerjee}}, \bibinfo {author} {\bibfnamefont {A.~M.}\ \bibnamefont
  {Ankowski}}, \bibinfo {author} {\bibfnamefont {K.~M.}\ \bibnamefont
  {Gra{\-}czyk}}, \bibinfo {author} {\bibfnamefont {B.~E.}\ \bibnamefont {Kowal}},
  \bibinfo {author} {\bibfnamefont {H.}~\bibnamefont {Prasad}},\ and\ \bibinfo
  {author} {\bibfnamefont {J.~T.}\ \bibnamefont {Sobczyk}},\ }\href
  {https://doi.org/10.1103/PhysRevD.109.073004} {\bibfield  {journal} {\bibinfo
   {journal} {Phys. Rev. D}\ }\textbf {\bibinfo {volume} {109}},\ \bibinfo
  {pages} {073004} (\bibinfo {year} {2024})}\BibitemShut {NoStop}%
\bibitem [{\citenamefont {Prasad}\ \emph {et~al.}(2025)\citenamefont {Prasad},
  \citenamefont {Sobczyk}, \citenamefont {Ankowski}, \citenamefont {Bonilla},
  \citenamefont {Dharmapal~Banerjee}, \citenamefont {Graczyk},\ and\
  \citenamefont {Kowal}}]{Prasad:2024gnv}%
  \BibitemOpen
  \bibfield  {author} {\bibinfo {author} {\bibfnamefont {H.}~\bibnamefont
  {Prasad}}, \bibinfo {author} {\bibfnamefont {J.~T.}\ \bibnamefont {Sobczyk}},
  \bibinfo {author} {\bibfnamefont {A.~M.}\ \bibnamefont {Ankowski}}, \bibinfo
  {author} {\bibfnamefont {J.~L.}\ \bibnamefont {Bonilla}}, \bibinfo {author}
  {\bibfnamefont {R.}~\bibnamefont {Dharmapal~Banerjee}}, \bibinfo {author}
  {\bibfnamefont {K.~M.}\ \bibnamefont {Graczyk}},\ and\ \bibinfo {author}
  {\bibfnamefont {B.~E.}\ \bibnamefont {Kowal}},\ }\href
  {https://doi.org/10.1103/PhysRevD.111.036032} {\bibfield  {journal} {\bibinfo
   {journal} {Phys. Rev. D}\ }\textbf {\bibinfo {volume} {111}},\ \bibinfo
  {pages} {036032} (\bibinfo {year} {2025})}\BibitemShut {NoStop}%
\bibitem [{\citenamefont {Jiang}\ \emph {et~al.}(2022)\citenamefont {Jiang}
  \emph {et~al.}}]{JeffersonLabHallA:2022cit}%
  \BibitemOpen
  \bibfield  {author} {\bibinfo {author} {\bibfnamefont {L.}~\bibnamefont
  {Jiang}} \emph {et~al.} (\bibinfo {collaboration} {Jefferson Lab Hall A
  Collaboration}),\ }\href {https://doi.org/10.1103/PhysRevD.105.112002}
  {\bibfield  {journal} {\bibinfo  {journal} {Phys. Rev. D}\ }\textbf {\bibinfo
  {volume} {105}},\ \bibinfo {pages} {112002} (\bibinfo {year}
  {2022})}\BibitemShut {NoStop}%
\bibitem [{\citenamefont {Jiang}\ \emph {et~al.}(2023)\citenamefont {Jiang}
  \emph {et~al.}}]{JeffersonLabHallA:2022ljj}%
  \BibitemOpen
  \bibfield  {author} {\bibinfo {author} {\bibfnamefont {L.}~\bibnamefont
  {Jiang}} \emph {et~al.} (\bibinfo {collaboration} {Jefferson Lab Hall A
  Collaboration}),\ }\href {https://doi.org/10.1103/PhysRevD.107.012005}
  {\bibfield  {journal} {\bibinfo  {journal} {Phys. Rev. D}\ }\textbf {\bibinfo
  {volume} {107}},\ \bibinfo {pages} {012005} (\bibinfo {year}
  {2023})}\BibitemShut {NoStop}%
\bibitem [{\citenamefont {Benhar}\ \emph {et~al.}(1994)\citenamefont {Benhar},
  \citenamefont {Fabrocini}, \citenamefont {Fantoni},\ and\ \citenamefont
  {Sick}}]{Benhar:1994hw}%
  \BibitemOpen
  \bibfield  {author} {\bibinfo {author} {\bibfnamefont {O.}~\bibnamefont
  {Benhar}}, \bibinfo {author} {\bibfnamefont {A.}~\bibnamefont {Fabrocini}},
  \bibinfo {author} {\bibfnamefont {S.}~\bibnamefont {Fantoni}},\ and\ \bibinfo
  {author} {\bibfnamefont {I.}~\bibnamefont {Sick}},\ }\href
  {https://doi.org/10.1016/0375-9474(94)90920-2} {\bibfield  {journal}
  {\bibinfo  {journal} {Nucl. Phys. A}\ }\textbf {\bibinfo {volume} {579}},\
  \bibinfo {pages} {493} (\bibinfo {year} {1994})}\BibitemShut {NoStop}%
\bibitem [{\citenamefont {Benhar}\ \emph {et~al.}(2005)\citenamefont {Benhar},
  \citenamefont {Farina}, \citenamefont {Nakamura}, \citenamefont {Sakuda},\
  and\ \citenamefont {Seki}}]{Benhar:2005dj}%
  \BibitemOpen
  \bibfield  {author} {\bibinfo {author} {\bibfnamefont {O.}~\bibnamefont
  {Benhar}}, \bibinfo {author} {\bibfnamefont {N.}~\bibnamefont {Farina}},
  \bibinfo {author} {\bibfnamefont {H.}~\bibnamefont {Nakamura}}, \bibinfo
  {author} {\bibfnamefont {M.}~\bibnamefont {Sakuda}},\ and\ \bibinfo {author}
  {\bibfnamefont {R.}~\bibnamefont {Seki}},\ }\href
  {https://doi.org/10.1103/PhysRevD.72.053005} {\bibfield  {journal} {\bibinfo
  {journal} {Phys. Rev. D}\ }\textbf {\bibinfo {volume} {72}},\ \bibinfo
  {pages} {053005} (\bibinfo {year} {2005})}\BibitemShut {NoStop}%
\bibitem [{\citenamefont {Mougey}\ \emph {et~al.}(1976)\citenamefont {Mougey},
  \citenamefont {Bernheim}, \citenamefont {Bussi{\`e}re}, \citenamefont
  {Gillibert}, \citenamefont {Xuan~H{\^o}}, \citenamefont {Priou},
  \citenamefont {Royer}, \citenamefont {Sick},\ and\ \citenamefont
  {Wagner}}]{Mougey:1976sc}%
  \BibitemOpen
  \bibfield  {author} {\bibinfo {author} {\bibfnamefont {J.}~\bibnamefont
  {Mougey}}, \bibinfo {author} {\bibfnamefont {M.}~\bibnamefont {Bernheim}},
  \bibinfo {author} {\bibfnamefont {A.}~\bibnamefont {Bussi{\`e}re}}, \bibinfo
  {author} {\bibfnamefont {A.}~\bibnamefont {Gillibert}}, \bibinfo {author}
  {\bibfnamefont {P.}~\bibnamefont {Xuan~H{\^o}}}, \bibinfo {author}
  {\bibfnamefont {M.}~\bibnamefont {Priou}}, \bibinfo {author} {\bibfnamefont
  {D.}~\bibnamefont {Royer}}, \bibinfo {author} {\bibfnamefont
  {I.}~\bibnamefont {Sick}},\ and\ \bibinfo {author} {\bibfnamefont {G.~J.}\
  \bibnamefont {Wagner}},\ }\href
  {https://doi.org/10.1016/0375-9474(76)90510-8} {\bibfield  {journal}
  {\bibinfo  {journal} {Nucl. Phys. A}\ }\textbf {\bibinfo {volume} {262}},\
  \bibinfo {pages} {461} (\bibinfo {year} {1976})}\BibitemShut {NoStop}%
\bibitem [{\citenamefont {Bernheim}\ \emph {et~al.}(1982)\citenamefont
  {Bernheim} \emph {et~al.}}]{Bernheim:1981si}%
  \BibitemOpen
  \bibfield  {author} {\bibinfo {author} {\bibfnamefont {M.}~\bibnamefont
  {Bernheim}} \emph {et~al.},\ }\href
  {https://doi.org/10.1016/0375-9474(82)90020-3} {\bibfield  {journal}
  {\bibinfo  {journal} {Nucl. Phys. A}\ }\textbf {\bibinfo {volume} {375}},\
  \bibinfo {pages} {381} (\bibinfo {year} {1982})}\BibitemShut {NoStop}%
\bibitem [{\citenamefont {Benhar}\ \emph {et~al.}(1989)\citenamefont {Benhar},
  \citenamefont {Fabrocini},\ and\ \citenamefont {Fantoni}}]{Benhar:1989aw}%
  \BibitemOpen
  \bibfield  {author} {\bibinfo {author} {\bibfnamefont {O.}~\bibnamefont
  {Benhar}}, \bibinfo {author} {\bibfnamefont {A.}~\bibnamefont {Fabrocini}},\
  and\ \bibinfo {author} {\bibfnamefont {S.}~\bibnamefont {Fantoni}},\ }\href
  {https://doi.org/10.1016/0375-9474(89)90374-6} {\bibfield  {journal}
  {\bibinfo  {journal} {Nucl. Phys. A}\ }\textbf {\bibinfo {volume} {505}},\
  \bibinfo {pages} {267} (\bibinfo {year} {1989})}\BibitemShut {NoStop}%
\bibitem [{\citenamefont {Ankowski}\ \emph {et~al.}(2024)\citenamefont
  {Ankowski}, \citenamefont {Benhar},\ and\ \citenamefont
  {Sakuda}}]{Ankowski:2024ntv}%
  \BibitemOpen
  \bibfield  {author} {\bibinfo {author} {\bibfnamefont {A.~M.}\ \bibnamefont
  {Ankowski}}, \bibinfo {author} {\bibfnamefont {O.}~\bibnamefont {Benhar}},\
  and\ \bibinfo {author} {\bibfnamefont {M.}~\bibnamefont {Sakuda}},\ }\href
  {https://doi.org/10.1103/PhysRevC.110.054612} {\bibfield  {journal} {\bibinfo
   {journal} {Phys. Rev. C}\ }\textbf {\bibinfo {volume} {110}},\ \bibinfo
  {pages} {054612} (\bibinfo {year} {2024})}\BibitemShut {NoStop}%
\bibitem [{\citenamefont {Van Der~Steenhoven}\ \emph
  {et~al.}(1988)\citenamefont {Van Der~Steenhoven}, \citenamefont {Blok},
  \citenamefont {Jans}, \citenamefont {De~Jong}, \citenamefont {Lapik\'as},
  \citenamefont {Quint},\ and\ \citenamefont
  {De~Witt~Huberts}}]{VanDerSteenhoven:1988qa}%
  \BibitemOpen
  \bibfield  {author} {\bibinfo {author} {\bibfnamefont {G.}~\bibnamefont {Van
  Der~Steenhoven}}, \bibinfo {author} {\bibfnamefont {H.~P.}\ \bibnamefont
  {Blok}}, \bibinfo {author} {\bibfnamefont {E.}~\bibnamefont {Jans}}, \bibinfo
  {author} {\bibfnamefont {M.}~\bibnamefont {De~Jong}}, \bibinfo {author}
  {\bibfnamefont {L.}~\bibnamefont {Lapik\'as}}, \bibinfo {author}
  {\bibfnamefont {E.~N.~M.}\ \bibnamefont {Quint}},\ and\ \bibinfo {author}
  {\bibfnamefont {P.~K.~A.}\ \bibnamefont {De~Witt~Huberts}},\ }\href
  {https://doi.org/10.1016/0375-9474(88)90463-0} {\bibfield  {journal}
  {\bibinfo  {journal} {Nucl. Phys. A}\ }\textbf {\bibinfo {volume} {480}},\
  \bibinfo {pages} {547} (\bibinfo {year} {1988})}\BibitemShut {NoStop}%
\bibitem [{\citenamefont {Sobczyk}\ \emph {et~al.}(2020)\citenamefont
  {Sobczyk}, \citenamefont {Nieves},\ and\ \citenamefont
  {S\'anchez}}]{Sobczyk:2020dkn}%
  \BibitemOpen
  \bibfield  {author} {\bibinfo {author} {\bibfnamefont {J.~E.}\ \bibnamefont
  {Sobczyk}}, \bibinfo {author} {\bibfnamefont {J.}~\bibnamefont {Nieves}},\
  and\ \bibinfo {author} {\bibfnamefont {F.}~\bibnamefont {S\'anchez}},\ }\href
  {https://doi.org/10.1103/PhysRevC.102.024601} {\bibfield  {journal} {\bibinfo
   {journal} {Phys. Rev. C}\ }\textbf {\bibinfo {volume} {102}},\ \bibinfo
  {pages} {024601} (\bibinfo {year} {2020})}\BibitemShut {NoStop}%
\bibitem [{\citenamefont {Yan}\ \emph {et~al.}()\citenamefont {Yan},
  \citenamefont {Niewczas}, \citenamefont {Nikolakopoulos}, \citenamefont
  {Gonz\'alez-Jim\'enez}, \citenamefont {Jachowicz}, \citenamefont {Lu},
  \citenamefont {Sobczyk},\ and\ \citenamefont {Zheng}}]{Yan:2024kkg}%
  \BibitemOpen
  \bibfield  {author} {\bibinfo {author} {\bibfnamefont {Q.}~\bibnamefont
  {Yan}}, \bibinfo {author} {\bibfnamefont {K.}~\bibnamefont {Niewczas}},
  \bibinfo {author} {\bibfnamefont {A.}~\bibnamefont {Nikolakopoulos}},
  \bibinfo {author} {\bibfnamefont {R.}~\bibnamefont {Gonz\'alez-Jim\'enez}},
  \bibinfo {author} {\bibfnamefont {N.}~\bibnamefont {Jachowicz}}, \bibinfo
  {author} {\bibfnamefont {X.}~\bibnamefont {Lu}}, \bibinfo {author}
  {\bibfnamefont {J.}~\bibnamefont {Sobczyk}},\ and\ \bibinfo {author}
  {\bibfnamefont {Y.}~\bibnamefont {Zheng}},\ }\href
  {https://doi.org/10.1007/JHEP12(2024)141} {\bibfield  {journal} {\bibinfo
  {journal} {J. High Energy Phys.}\ }\textbf {\bibinfo {volume} {12}}\bibinfo
  {number} { (2024)}\ \bibinfo {pages} {141}}\BibitemShut {NoStop}%
\bibitem [{\citenamefont {Benhar}\ \emph {et~al.}(1991)\citenamefont {Benhar},
  \citenamefont {Fabrocini}, \citenamefont {Fantoni}, \citenamefont {Miller},
  \citenamefont {Pandharipande},\ and\ \citenamefont {Sick}}]{Benhar:1991af}%
  \BibitemOpen
\bibfield  {number} {  }\bibfield  {author} {\bibinfo {author} {\bibfnamefont
  {O.}~\bibnamefont {Benhar}}, \bibinfo {author} {\bibfnamefont
  {A.}~\bibnamefont {Fabrocini}}, \bibinfo {author} {\bibfnamefont
  {S.}~\bibnamefont {Fantoni}}, \bibinfo {author} {\bibfnamefont {G.~A.}\
  \bibnamefont {Miller}}, \bibinfo {author} {\bibfnamefont {V.~R.}\
  \bibnamefont {Pandharipande}},\ and\ \bibinfo {author} {\bibfnamefont
  {I.}~\bibnamefont {Sick}},\ }\href {https://doi.org/10.1103/PhysRevC.44.2328}
  {\bibfield  {journal} {\bibinfo  {journal} {Phys. Rev. C}\ }\textbf {\bibinfo
  {volume} {44}},\ \bibinfo {pages} {2328} (\bibinfo {year}
  {1991})}\BibitemShut {NoStop}%
\bibitem [{\citenamefont {Benhar}\ \emph {et~al.}(2008)\citenamefont {Benhar},
  \citenamefont {Day},\ and\ \citenamefont {Sick}}]{Benhar:2006wy}%
  \BibitemOpen
  \bibfield  {author} {\bibinfo {author} {\bibfnamefont {O.}~\bibnamefont
  {Benhar}}, \bibinfo {author} {\bibfnamefont {D.}~\bibnamefont {Day}},\ and\
  \bibinfo {author} {\bibfnamefont {I.}~\bibnamefont {Sick}},\ }\href
  {https://doi.org/10.1103/RevModPhys.80.189} {\bibfield  {journal} {\bibinfo
  {journal} {Rev. Mod. Phys.}\ }\textbf {\bibinfo {volume} {80}},\ \bibinfo
  {pages} {189} (\bibinfo {year} {2008})}\BibitemShut {NoStop}%
\bibitem [{\citenamefont {Benhar}(2013)}]{Benhar:2013dq}%
  \BibitemOpen
  \bibfield  {author} {\bibinfo {author} {\bibfnamefont {O.}~\bibnamefont
  {Benhar}},\ }\href {https://doi.org/10.1103/PhysRevC.87.024606} {\bibfield
  {journal} {\bibinfo  {journal} {Phys. Rev. C}\ }\textbf {\bibinfo {volume}
  {87}},\ \bibinfo {pages} {024606} (\bibinfo {year} {2013})}\BibitemShut
  {NoStop}%
\bibitem [{\citenamefont {Ankowski}\ \emph {et~al.}(2015)\citenamefont
  {Ankowski}, \citenamefont {Benhar},\ and\ \citenamefont
  {Sakuda}}]{Ankowski:2014yfa}%
  \BibitemOpen
  \bibfield  {author} {\bibinfo {author} {\bibfnamefont {A.~M.}\ \bibnamefont
  {Ankowski}}, \bibinfo {author} {\bibfnamefont {O.}~\bibnamefont {Benhar}},\
  and\ \bibinfo {author} {\bibfnamefont {M.}~\bibnamefont {Sakuda}},\ }\href
  {https://doi.org/10.1103/PhysRevD.91.033005} {\bibfield  {journal} {\bibinfo
  {journal} {Phys. Rev. D}\ }\textbf {\bibinfo {volume} {91}},\ \bibinfo
  {pages} {033005} (\bibinfo {year} {2015})}\BibitemShut {NoStop}%
\bibitem [{\citenamefont {Golan}\ \emph {et~al.}(2012)\citenamefont {Golan},
  \citenamefont {Juszczak},\ and\ \citenamefont {Sobczyk}}]{Golan:2012wx}%
  \BibitemOpen
  \bibfield  {author} {\bibinfo {author} {\bibfnamefont {T.}~\bibnamefont
  {Golan}}, \bibinfo {author} {\bibfnamefont {C.}~\bibnamefont {Juszczak}},\
  and\ \bibinfo {author} {\bibfnamefont {J.~T.}\ \bibnamefont {Sobczyk}},\
  }\href {https://doi.org/10.1103/PhysRevC.86.015505} {\bibfield  {journal}
  {\bibinfo  {journal} {Phys. Rev. C}\ }\textbf {\bibinfo {volume} {86}},\
  \bibinfo {pages} {015505} (\bibinfo {year} {2012})}\BibitemShut {NoStop}%
\bibitem [{\citenamefont {Niewczas}\ and\ \citenamefont
  {Sobczyk}(2019)}]{Niewczas:2019fro}%
  \BibitemOpen
  \bibfield  {author} {\bibinfo {author} {\bibfnamefont {K.}~\bibnamefont
  {Niewczas}}\ and\ \bibinfo {author} {\bibfnamefont {J.~T.}\ \bibnamefont
  {Sobczyk}},\ }\href {https://doi.org/10.1103/PhysRevC.100.015505} {\bibfield
  {journal} {\bibinfo  {journal} {Phys. Rev. C}\ }\textbf {\bibinfo {volume}
  {100}},\ \bibinfo {pages} {015505} (\bibinfo {year} {2019})}\BibitemShut
  {NoStop}%
\bibitem [{\citenamefont {Dytman}\ \emph {et~al.}(2021)\citenamefont {Dytman},
  \citenamefont {Hayato}, \citenamefont {Raboanary}, \citenamefont {Sobczyk},
  \citenamefont {Tena~Vidal},\ and\ \citenamefont
  {Vololoniaina}}]{Dytman:2021ohr}%
  \BibitemOpen
  \bibfield  {author} {\bibinfo {author} {\bibfnamefont {S.}~\bibnamefont
  {Dytman}}, \bibinfo {author} {\bibfnamefont {Y.}~\bibnamefont {Hayato}},
  \bibinfo {author} {\bibfnamefont {R.}~\bibnamefont {Raboanary}}, \bibinfo
  {author} {\bibfnamefont {J.~T.}\ \bibnamefont {Sobczyk}}, \bibinfo {author}
  {\bibfnamefont {J.}~\bibnamefont {Tena~Vidal}},\ and\ \bibinfo {author}
  {\bibfnamefont {N.}~\bibnamefont {Vololoniaina}},\ }\href
  {https://doi.org/10.1103/PhysRevD.104.053006} {\bibfield  {journal} {\bibinfo
   {journal} {Phys. Rev. D}\ }\textbf {\bibinfo {volume} {104}},\ \bibinfo
  {pages} {053006} (\bibinfo {year} {2021})}\BibitemShut {NoStop}%
\bibitem [{\citenamefont {De~Vries}\ \emph {et~al.}(1987)\citenamefont
  {De~Vries}, \citenamefont {De~Jager},\ and\ \citenamefont
  {De~Vries}}]{DeVries:1987atn}%
  \BibitemOpen
  \bibfield  {author} {\bibinfo {author} {\bibfnamefont {H.}~\bibnamefont
  {De~Vries}}, \bibinfo {author} {\bibfnamefont {C.~W.}\ \bibnamefont
  {De~Jager}},\ and\ \bibinfo {author} {\bibfnamefont {C.}~\bibnamefont
  {De~Vries}},\ }\href {https://doi.org/10.1016/0092-640X(87)90013-1}
  {\bibfield  {journal} {\bibinfo  {journal} {Atom. Data Nucl. Data Tabl.}\
  }\textbf {\bibinfo {volume} {36}},\ \bibinfo {pages} {495} (\bibinfo {year}
  {1987})}\BibitemShut {NoStop}%
\bibitem [{\citenamefont {Kelly}(2002)}]{Kelly:2002if}%
  \BibitemOpen
  \bibfield  {author} {\bibinfo {author} {\bibfnamefont {J.~J.}\ \bibnamefont
  {Kelly}},\ }\href {https://doi.org/10.1103/PhysRevC.66.065203} {\bibfield
  {journal} {\bibinfo  {journal} {Phys. Rev. C}\ }\textbf {\bibinfo {volume}
  {66}},\ \bibinfo {pages} {065203} (\bibinfo {year} {2002})}\BibitemShut
  {NoStop}%
\bibitem [{\citenamefont {Benhar}()}]{Benhar:2003ka}%
  \BibitemOpen
  \bibfield  {author} {\bibinfo {author} {\bibfnamefont {O.}~\bibnamefont
  {Benhar}},\ }\Eprint {https://arxiv.org/abs/nucl-th/0307061}
  {arXiv:nucl-th/0307061} (\bibinfo {year} {2003})\BibitemShut {NoStop}%
\bibitem [{\citenamefont {Rohe}\ \emph {et~al.}(2005)\citenamefont {Rohe} \emph
  {et~al.}}]{E97-006:2005jlg}%
  \BibitemOpen
  \bibfield  {author} {\bibinfo {author} {\bibfnamefont {D.}~\bibnamefont
  {Rohe}} \emph {et~al.} (\bibinfo {collaboration} {E97-006 Collaboration}),\
  }\href {https://doi.org/10.1103/PhysRevC.72.054602} {\bibfield  {journal}
  {\bibinfo  {journal} {Phys. Rev. C}\ }\textbf {\bibinfo {volume} {72}},\
  \bibinfo {pages} {054602} (\bibinfo {year} {2005})}\BibitemShut {NoStop}%
\bibitem [{\citenamefont {Garino}\ \emph {et~al.}(1992)\citenamefont {Garino}
  \emph {et~al.}}]{Garino:1992ca}%
  \BibitemOpen
  \bibfield  {author} {\bibinfo {author} {\bibfnamefont {G.}~\bibnamefont
  {Garino}} \emph {et~al.},\ }\href {https://doi.org/10.1103/PhysRevC.45.780}
  {\bibfield  {journal} {\bibinfo  {journal} {Phys. Rev. C}\ }\textbf {\bibinfo
  {volume} {45}},\ \bibinfo {pages} {780} (\bibinfo {year} {1992})}\BibitemShut
  {NoStop}%
\bibitem [{\citenamefont {O'Neill}\ \emph {et~al.}(1995)\citenamefont {O'Neill}
  \emph {et~al.}}]{ONeill:1994znv}%
  \BibitemOpen
  \bibfield  {author} {\bibinfo {author} {\bibfnamefont {T.~G.}\ \bibnamefont
  {O'Neill}} \emph {et~al.},\ }\href
  {https://doi.org/10.1016/0370-2693(95)00362-O} {\bibfield  {journal}
  {\bibinfo  {journal} {Phys. Lett. B}\ }\textbf {\bibinfo {volume} {351}},\
  \bibinfo {pages} {87} (\bibinfo {year} {1995})}\BibitemShut {NoStop}%
\bibitem [{\citenamefont {Garrow}\ \emph {et~al.}(2002)\citenamefont {Garrow}
  \emph {et~al.}}]{Garrow:2001di}%
  \BibitemOpen
  \bibfield  {author} {\bibinfo {author} {\bibfnamefont {K.}~\bibnamefont
  {Garrow}} \emph {et~al.},\ }\href
  {https://doi.org/10.1103/PhysRevC.66.044613} {\bibfield  {journal} {\bibinfo
  {journal} {Phys. Rev. C}\ }\textbf {\bibinfo {volume} {66}},\ \bibinfo
  {pages} {044613} (\bibinfo {year} {2002})}\BibitemShut {NoStop}%
\bibitem [{\citenamefont {Dutta}\ \emph {et~al.}(2003)\citenamefont {Dutta}
  \emph {et~al.}}]{JLabE91013:2003gdp}%
  \BibitemOpen
  \bibfield  {author} {\bibinfo {author} {\bibfnamefont {D.}~\bibnamefont
  {Dutta}} \emph {et~al.},\ }\href {https://doi.org/10.1103/PhysRevC.68.064603}
  {\bibfield  {journal} {\bibinfo  {journal} {Phys. Rev. C}\ }\textbf {\bibinfo
  {volume} {68}},\ \bibinfo {pages} {064603} (\bibinfo {year}
  {2003})}\BibitemShut {NoStop}%
\bibitem [{\citenamefont {Bhetuwal}\ \emph {et~al.}(2023)\citenamefont
  {Bhetuwal} \emph {et~al.}}]{HallC:2022qlb}%
  \BibitemOpen
  \bibfield  {author} {\bibinfo {author} {\bibfnamefont {D.}~\bibnamefont
  {Bhetuwal}} \emph {et~al.} (\bibinfo {collaboration} {Hall C
  Collaboration}),\ }\href {https://doi.org/10.1103/PhysRevC.108.025203}
  {\bibfield  {journal} {\bibinfo  {journal} {Phys. Rev. C}\ }\textbf {\bibinfo
  {volume} {108}},\ \bibinfo {pages} {025203} (\bibinfo {year}
  {2023})}\BibitemShut {NoStop}%
\bibitem [{\citenamefont {Cooper}\ \emph {et~al.}(1993)\citenamefont {Cooper},
  \citenamefont {Hama}, \citenamefont {Clark},\ and\ \citenamefont
  {Mercer}}]{Cooper:1993nx}%
  \BibitemOpen
  \bibfield  {author} {\bibinfo {author} {\bibfnamefont {E.~D.}\ \bibnamefont
  {Cooper}}, \bibinfo {author} {\bibfnamefont {S.}~\bibnamefont {Hama}},
  \bibinfo {author} {\bibfnamefont {B.~C.}\ \bibnamefont {Clark}},\ and\
  \bibinfo {author} {\bibfnamefont {R.~L.}\ \bibnamefont {Mercer}},\ }\href
  {https://doi.org/10.1103/PhysRevC.47.297} {\bibfield  {journal} {\bibinfo
  {journal} {Phys. Rev. C}\ }\textbf {\bibinfo {volume} {47}},\ \bibinfo
  {pages} {297} (\bibinfo {year} {1993})}\BibitemShut {NoStop}%
\bibitem [{\citenamefont {Ciofi~degli Atti}\ and\ \citenamefont
  {Simula}(1996)}]{CiofidegliAtti:1995qe}%
  \BibitemOpen
  \bibfield  {author} {\bibinfo {author} {\bibfnamefont {C.}~\bibnamefont
  {Ciofi~degli Atti}}\ and\ \bibinfo {author} {\bibfnamefont {S.}~\bibnamefont
  {Simula}},\ }\href {https://doi.org/10.1103/PhysRevC.53.1689} {\bibfield
  {journal} {\bibinfo  {journal} {Phys. Rev. C}\ }\textbf {\bibinfo {volume}
  {53}},\ \bibinfo {pages} {1689} (\bibinfo {year} {1996})}\BibitemShut
  {NoStop}%
\bibitem [{\citenamefont {Nguyen}\ \emph {et~al.}(2020)\citenamefont {Nguyen}
  \emph {et~al.}}]{JeffersonLabHallA:2020wrr}%
  \BibitemOpen
  \bibfield  {author} {\bibinfo {author} {\bibfnamefont {D.}~\bibnamefont
  {Nguyen}} \emph {et~al.} (\bibinfo {collaboration} {Jefferson Lab Hall A
  Collaboration}),\ }\href {https://doi.org/10.1103/PhysRevC.102.064004}
  {\bibfield  {journal} {\bibinfo  {journal} {Phys. Rev. C}\ }\textbf {\bibinfo
  {volume} {102}},\ \bibinfo {pages} {064004} (\bibinfo {year}
  {2020})}\BibitemShut {NoStop}%
\bibitem [{\citenamefont {Barreau}\ \emph {et~al.}(1983)\citenamefont {Barreau}
  \emph {et~al.}}]{Barreau:1983ht}%
  \BibitemOpen
  \bibfield  {author} {\bibinfo {author} {\bibfnamefont {P.}~\bibnamefont
  {Barreau}} \emph {et~al.},\ }\href
  {https://doi.org/10.1016/0375-9474(83)90217-8} {\bibfield  {journal}
  {\bibinfo  {journal} {Nucl. Phys.}\ }\textbf {\bibinfo {volume} {A402}},\
  \bibinfo {pages} {515} (\bibinfo {year} {1983})}\BibitemShut {NoStop}%
\bibitem [{\citenamefont {Sealock}\ \emph {et~al.}(1989)\citenamefont {Sealock}
  \emph {et~al.}}]{Sealock:1989nx}%
  \BibitemOpen
  \bibfield  {author} {\bibinfo {author} {\bibfnamefont {R.~M.}\ \bibnamefont
  {Sealock}} \emph {et~al.},\ }\href
  {https://doi.org/10.1103/PhysRevLett.62.1350} {\bibfield  {journal} {\bibinfo
   {journal} {Phys. Rev. Lett.}\ }\textbf {\bibinfo {volume} {62}},\ \bibinfo
  {pages} {1350} (\bibinfo {year} {1989})}\BibitemShut {NoStop}%
\bibitem [{\citenamefont {Ankowski}\ and\ \citenamefont
  {Friedland}(2020)}]{Ankowski:2020qbe}%
  \BibitemOpen
  \bibfield  {author} {\bibinfo {author} {\bibfnamefont {A.~M.}\ \bibnamefont
  {Ankowski}}\ and\ \bibinfo {author} {\bibfnamefont {A.}~\bibnamefont
  {Friedland}},\ }\href {https://doi.org/10.1103/PhysRevD.102.053001}
  {\bibfield  {journal} {\bibinfo  {journal} {Phys. Rev. D}\ }\textbf {\bibinfo
  {volume} {102}},\ \bibinfo {pages} {053001} (\bibinfo {year}
  {2020})}\BibitemShut {NoStop}%
\bibitem [{\citenamefont {Kowal}\ \emph {et~al.}(2024)\citenamefont {Kowal},
  \citenamefont {Graczyk}, \citenamefont {Ankowski}, \citenamefont {Banerjee},
  \citenamefont {Prasad},\ and\ \citenamefont {Sobczyk}}]{Kowal:2023dcq}%
  \BibitemOpen
  \bibfield  {author} {\bibinfo {author} {\bibfnamefont {B.~E.}\ \bibnamefont
  {Kowal}}, \bibinfo {author} {\bibfnamefont {K.~M.}\ \bibnamefont {Graczyk}},
  \bibinfo {author} {\bibfnamefont {A.~M.}\ \bibnamefont {Ankowski}}, \bibinfo
  {author} {\bibfnamefont {R.~D.}\ \bibnamefont {Banerjee}}, \bibinfo {author}
  {\bibfnamefont {H.}~\bibnamefont {Prasad}},\ and\ \bibinfo {author}
  {\bibfnamefont {J.~T.}\ \bibnamefont {Sobczyk}},\ }\href
  {https://doi.org/10.1103/PhysRevC.110.025501} {\bibfield  {journal} {\bibinfo
   {journal} {Phys. Rev. C}\ }\textbf {\bibinfo {volume} {110}},\ \bibinfo
  {pages} {025501} (\bibinfo {year} {2024})}\BibitemShut {NoStop}%
\bibitem [{\citenamefont {Ankowski}\ \emph {et~al.}(2023)\citenamefont
  {Ankowski} \emph {et~al.}}]{Ankowski:2022thw}%
  \BibitemOpen
  \bibfield  {author} {\bibinfo {author} {\bibfnamefont {A.~M.}\ \bibnamefont
  {Ankowski}} \emph {et~al.},\ }\href
  {https://doi.org/10.1088/1361-6471/acef42} {\bibfield  {journal} {\bibinfo
  {journal} {J. Phys. G}\ }\textbf {\bibinfo {volume} {50}},\ \bibinfo {pages}
  {120501} (\bibinfo {year} {2023})}\BibitemShut {NoStop}%
\bibitem [{\citenamefont {Lu}\ \emph {et~al.}(2016)\citenamefont {Lu},
  \citenamefont {Pickering}, \citenamefont {Dolan}, \citenamefont {Barr},
  \citenamefont {Coplowe}, \citenamefont {Uchida}, \citenamefont {Wark},
  \citenamefont {Wascko}, \citenamefont {Weber},\ and\ \citenamefont
  {Yuan}}]{Lu:2015tcr}%
  \BibitemOpen
  \bibfield  {author} {\bibinfo {author} {\bibfnamefont {X.~G.}\ \bibnamefont
  {Lu}}, \bibinfo {author} {\bibfnamefont {L.}~\bibnamefont {Pickering}},
  \bibinfo {author} {\bibfnamefont {S.}~\bibnamefont {Dolan}}, \bibinfo
  {author} {\bibfnamefont {G.}~\bibnamefont {Barr}}, \bibinfo {author}
  {\bibfnamefont {D.}~\bibnamefont {Coplowe}}, \bibinfo {author} {\bibfnamefont
  {Y.}~\bibnamefont {Uchida}}, \bibinfo {author} {\bibfnamefont
  {D.}~\bibnamefont {Wark}}, \bibinfo {author} {\bibfnamefont {M.~O.}\
  \bibnamefont {Wascko}}, \bibinfo {author} {\bibfnamefont {A.}~\bibnamefont
  {Weber}},\ and\ \bibinfo {author} {\bibfnamefont {T.}~\bibnamefont {Yuan}},\
  }\href {https://doi.org/10.1103/PhysRevC.94.015503} {\bibfield  {journal}
  {\bibinfo  {journal} {Phys. Rev. C}\ }\textbf {\bibinfo {volume} {94}},\
  \bibinfo {pages} {015503} (\bibinfo {year} {2016})}\BibitemShut {NoStop}%
\bibitem [{Sup()}]{SupplemementalMaterial}%
  \BibitemOpen
  \href@noop {} {}\bibinfo {howpublished} {See Supplemental Material for
  comparisons to the CC$1p0\pi$ double differential cross
  sections.}\BibitemShut {Stop}%
\end{thebibliography}

%

\end{document}